\documentclass[amsmath,amssymb,aps,prc,reprint,superscriptaddress]{revtex4-1}
\usepackage{booktabs}
\usepackage{graphicx}
\usepackage{subfigure}
\usepackage{float}
\usepackage{xcolor}
\usepackage{dcolumn}



\begin{document}
\preprint{APS/123-QED}
\title{Evidence for the $^{15}\text{Be}$ ground state from $^{12}$Be$+3$n events}



\author{A.~N.~Kuchera}
\email[]{Corresponding author: ankuchera@davidson.edu}
\affiliation{Department of Physics, Davidson College, Davidson, North Carolina 28035, USA}

\author{R.~Shahid}
\affiliation{Department of Physics, Davidson College, Davidson, North Carolina 28035, USA}

\author{J.~Zhao}
\affiliation{Department of Physics, Davidson College, Davidson, North Carolina 28035, USA}

\author{A.~Edmondson}
\affiliation{Department of Physics, Davidson College, Davidson, North Carolina 28035, USA}

\author{P.~A.~DeYoung}
\affiliation{Department of Physics, Hope College, Holland, Michigan 49423, USA}

\author{N. Frank}
\affiliation{Department of Physics, Augustana College, Rock Island, Illinois 61201, USA}

\author{J. McDonaugh}
\affiliation{Department of Physics, Augustana College, Rock Island, Illinois 61201, USA}

\author{O. Peterson-Veatch}
\affiliation{Department of Physics, Augustana College, Rock Island, Illinois 61201, USA}

\author{W.~F.~Rogers}
\affiliation{Department of Physics, Indiana Wesleyan University, Marion, Indiana 46953}

\author{T.~Redpath}
\affiliation{Department of Chemistry, Virginia State University, Petersburg, Virginia 23806, USA}

\author{M.~Thoennessen}
\affiliation{National Superconducting Cyclotron Laboratory, Michigan State University, East Lansing, Michigan 48824, USA}
\affiliation{Department of Physics and Astronomy, Michigan State University, East Lansing, Michigan 48824, USA}

\begin{abstract}
\textbf{Background:} $^{15}$Be is an unbound nuclide that has been observed to decay by one-neutron emission. Shell model calculations predict two low-lying states in its energy spectrum, however, only a single resonance has been observed from coincident measurements of $^{14}$Be+n. It has been suggested that the yet unobserved state may decay sequentially through the first excited state in $^{14}$Be followed by a two-neutron emission to $^{12}$Be.

\textbf{Purpose:} The ground state of $^{15}$Be has yet to be confirmed. A search for this predicted $^{15}$Be state by reconstructing $^{12}$Be$+3$n events allows a possible determination of its ground state properties.

\textbf{Methods:} A neutron-pickup reaction was performed with a $^{14}$Be beam on a CD$_2$ target to populate unbound $^{15}$Be states. Decay energies were reconstructed using invariant mass spectroscopy by detecting $^{12}$Be daughter nuclei in coincidence with up to three neutrons.

\textbf{Results:} Evidence for at least one resonance in $^{15}$Be is presented based on the reconstruction of $^{12}$Be$+3$n events. Through comparison with simulations, the energy of the strongest resonance in the analyzed reaction and decay channel is determined to be $E_{^{12}Be+3n}=330(20)$ keV.

\textbf{Conclusions:} The inclusion of a new $^{15}$Be state among the $^{12}$Be+$3$n events lower in relative decay energy than the previous $^{14}$Be+n observations provides the best fit to the data. Because this suggested new state would be lower in energy than the previously observed state, it is a candidate for the ground state of $^{15}$Be.

\end{abstract}

\maketitle


\section{Introduction}
The exploration of drip-line nuclei is an important way to study the limits of nuclear structure. Neutron-rich beryllium isotopes demonstrate a variety of interesting structural and decay phenomena such as the direct emission of two neutrons from $^{16}$Be \cite{Spyrou2012, Marques2012, Lovell2017, Casal2018, Casal2019, Fortune2019, Belen2024}. To understand the details of how $^{16}$Be decays requires measurements of the level structure of $^{15}$Be to rule out the possibility of a sequential decay to $^{14}$Be. 

\begin{figure}[htbp] 
   \includegraphics[width=8.6cm]{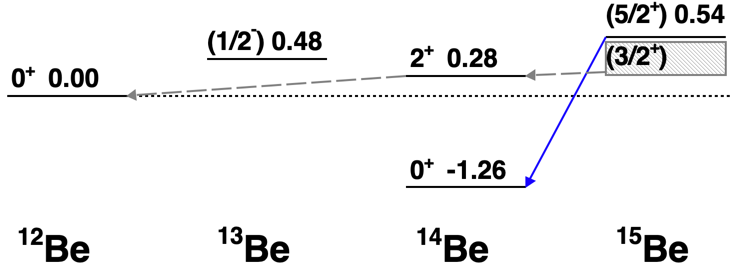} 
   \caption{The solid blue arrow represents the decay path of the previously observed $^{15}$Be state via $^{14}$Be$+$n events. States in $^{15}$Be with energies greater than the first excited state of $^{14}$Be could decay sequentially through it to $^{12}$Be$+3$n. This region, shown by the gray box with dashed gray lines, is the energy range the $3/2^+$ state is predicted to exist within. Energies are listed in MeV.}
   \label{fig:old_levels}
\end{figure}

The first attempt to observe $^{15}$Be used a two-proton removal reaction from a $^{17}$C beam and searched for $^{14}$Be+n coincidences with the MoNA+Sweeper setup at the National Superconducting Cyclotron Laboratory (NSCL) \cite{Spyrou2011}. This reaction is expected to populate the $3/2^{+}$ state as it has the same neutron configuration as the ground state of the $^{17}$C beam. The authors performed shell model calculations using NuShellX with the WBP Hamilton and predicted a $3/2^+$ ground state with a nearby $5/2^+$ state. However, a non-observation was reported owing to the lack of coincident events \cite{Spyrou2011}. Spectroscopic factors from the shell model calculations were determined for the $^{15}$Be $3/2^+$ state to the $2^+$ state in $^{14}$Be to be 1.27 and 0.084 for $\ell=2$ and $\ell=0$ orbital angular momentum transfers, respectively. It was reported that the non-observation of $^{14}$Be+n events may be due to the $3/2^+$ state decaying sequentially through the first excited state in $^{14}$Be, which is neutron unbound. Because the lowest observed $^{13}$Be state is energetically above the $^{14}$Be*(2$^+$) state, the $^{15}$Be state decaying through this state would then proceed to the next bound isotope, $^{12}$Be, by the emission of two more neutrons. See Fig.~\ref{fig:old_levels} for a level diagram and suggested decay path.

A second experiment at NSCL with the MoNA+Sweeper setup used a neutron-pickup reaction on a CD$_2$ target. A resonance was observed decaying to the ground state of $^{14}$Be with $E_{^{14}Be+n}=1.8(1)$ MeV and the authors suggested $J^{\pi}=5/2^+$ \cite{Snyder}. This was the first observation of $^{15}$Be. The resonance was recently confirmed by a multiple-nucleon knockout reaction from $^{18}$C at RIKEN using SAMURAI+NEBULA \cite{Belen}. A compatible energy of 1.70$\pm$0.13 MeV was reported from the fitting of this resonance in addition to a background and contributions from the $^{16}$Be ground state. While three low-lying states have been predicted, only one resonance was observed in each  of the two experiments. As suggested in Ref. \cite{Spyrou2011}, an alternative decay path might explain the non-observation of the other predicted states.

The first attempt to search for three-neutron events from $^{15}$Be was performed by Kuchera \textit{et al.} \cite{Kuchera} using the data from Spyrou \textit{et al.} \cite{Spyrou2011}. The two-, three-, or four-body decay energies were simultaneously fitted using $^{12}$Be fragments and one, two, and three neutrons, respectively. Previously observed resonances in $^{13}$Be and $^{14}$Be were fixed and the simulation of a $^{15}$Be state was included with its energy as a free parameter. Owing to low statistics, only a broad range of possibilities could be suggested. However, there was not convincing evidence for a state in $^{15}$Be being required to describe the data. 

While previous calculations predicted a small overlap for the $5/2^+$ state with the $2^+$ state in $^{14}$Be, Fortune computed widths and spectroscopic factors indicating a significant decay strength through this channel \cite{Fortune2015, Fortune2018}. The strong support for $^{15}$Be decays through $^{14}$Be*($2^+$) motivated a re-analysis of the data from Snyder \textit{et al.} \cite{Snyder} and is the focus of this work.

\section{Experiment}

The experiment was performed at the Coupled Cyclotron Facility at the NSCL. A 59 MeV/nucleon $^{14}$Be beam was produced from a 120 MeV/nucleon $^{18}$O primary beam by projectile fragmentation on a Be target in the A1900 fragment separator. The $^{14}$Be beam was focused onto a solid 435 mg/cm$^2$ CD$_2$ target. Charged particles downstream of the target were bent $43^{\circ}$ by the large-gap dipole Sweeper magnet into a suite of charged particle detectors \cite{sweeper}. The emitted neutrons continued in the beam direction toward the Modular Neutron Array (MoNA) \cite{mona1, mona2}. 

\begin{figure}[htbp] 
   \includegraphics[width=8.6cm]{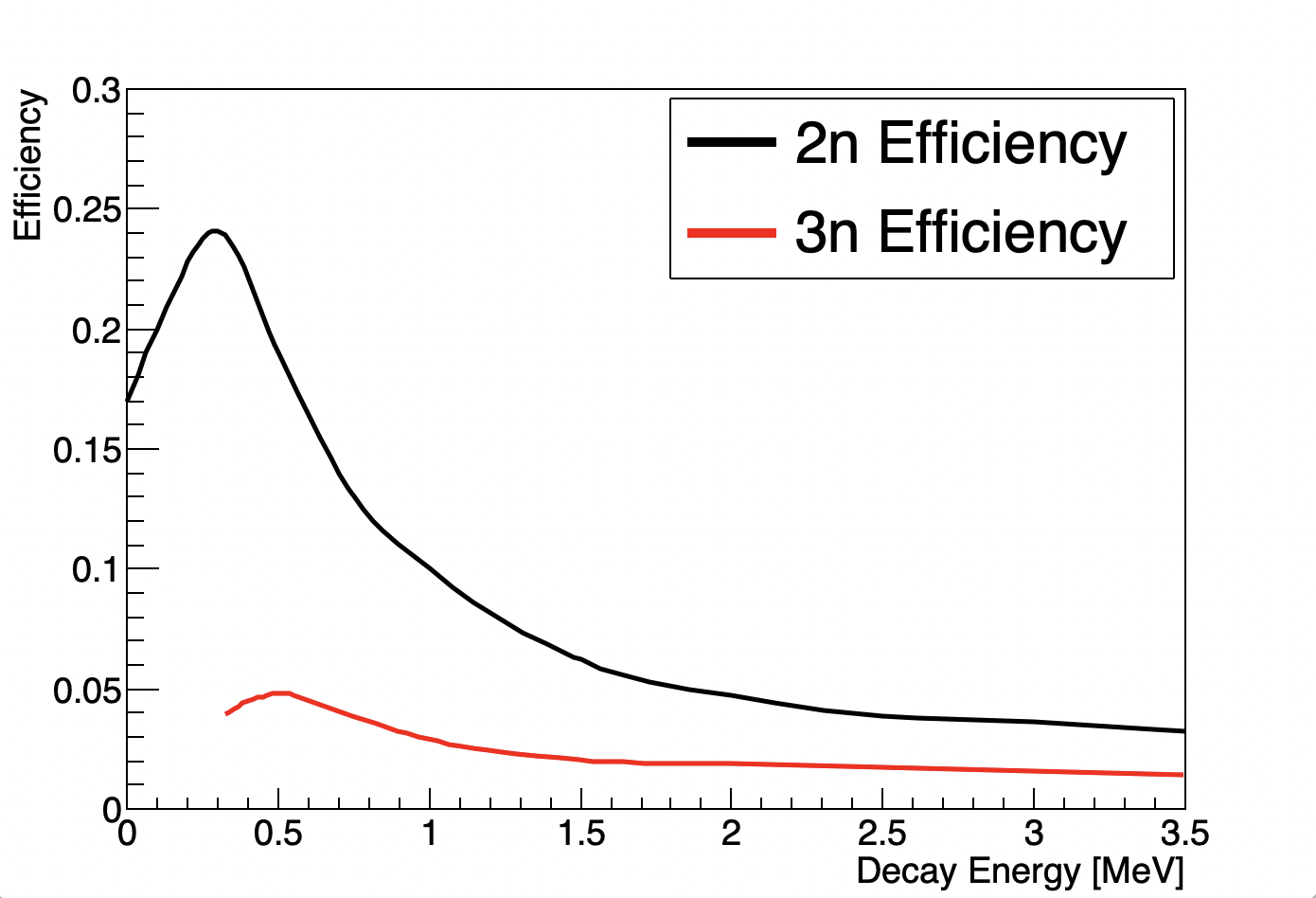} 
   \caption{Three-body (black) and four-body (red) detection efficiencies for the experimental setup as a function of decay energy.}
   \label{fig:efficiency}
\end{figure}

\begin{figure}[htbp] 
   \includegraphics[width=8.6cm]{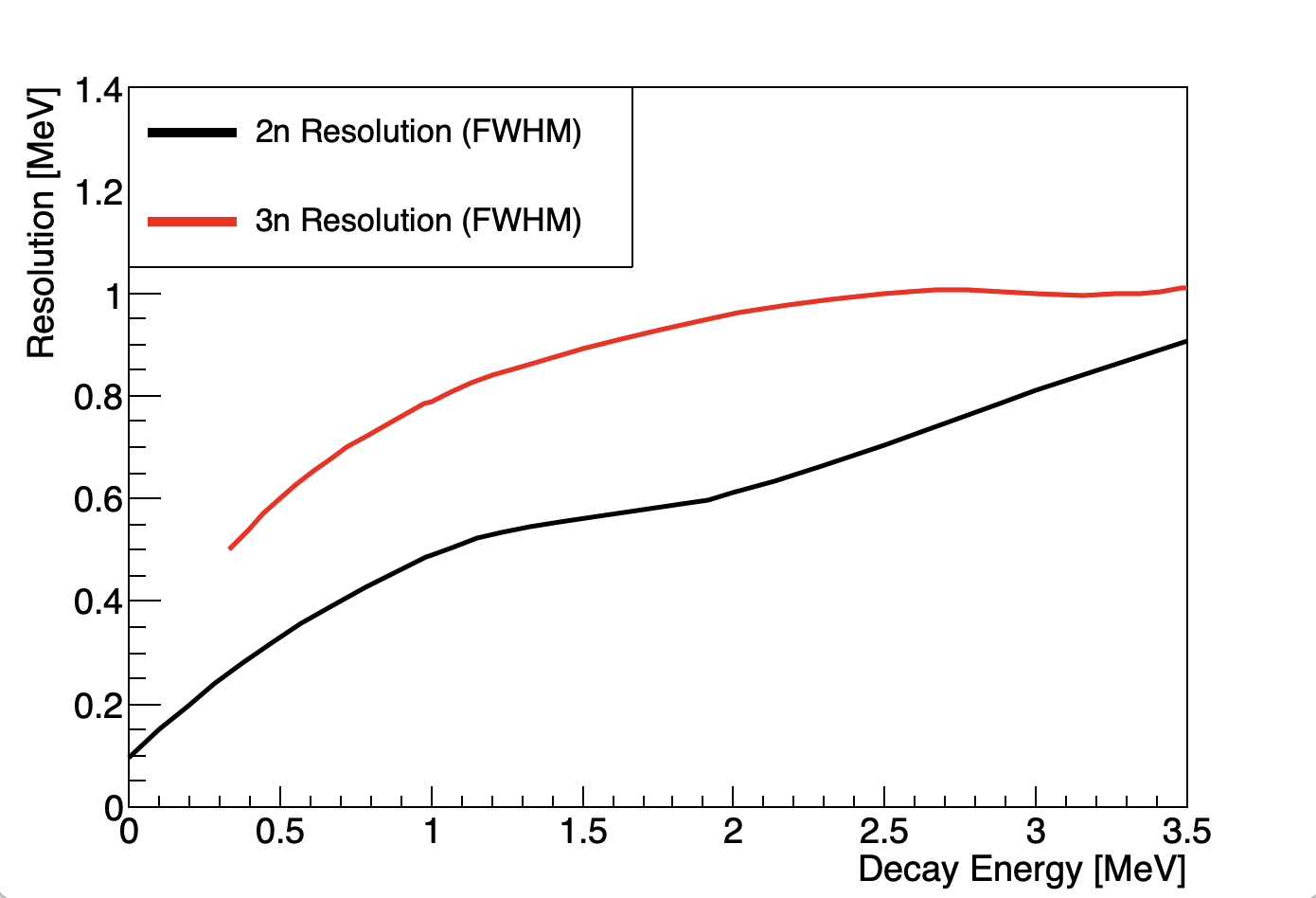} 
   \caption{Three-body (black) and four-body (red) decay energy resolutions in full width half maximum for the experimental setup as a function of decay energy.}
   \label{fig:resolution}
\end{figure}

Beryllium events were identified based on their energy loss in an ionization chamber. The $^{12}$Be isotope was selected based on its corrected time of flight from the reaction target to a large area timing scintillator at the end of the focal plane detector suite. Invariant mass spectroscopy was used to reconstruct the decay energies of the unbound states of nuclei that were produced. The first three time-ordered hits in MoNA in coincidence with $^{12}$Be were used to reconstruct four-body decay energies. The detection efficiency for three-body ($^{12}$Be+2n) and four-body ($^{12}$Be+3n) decays are shown in Fig~\ref{fig:efficiency}. The 3n efficiency in the decay energy range of interest is approximately 1-5\%. The decay energy resolution (full width half maximum) as a function of decay energy for three-body ($^{12}$Be+2n) and four-body ($^{12}$Be+3n) decays are shown in Fig~\ref{fig:resolution}. In both cases the values are determined from Monte Carlo simulations of the experiment and causality cuts were applied (discussed in Section III). The inclusion of causality cuts reduces the efficiency at lower decay energies and the resolution is mainly affected by target thickness and neutron detection position resolution. It should be noted that the 3n values are only shown down to the decay energy values explored in this analysis and are not an indication of the 3n resolution and efficiency limits the experimental setup is capable of detecting. Because the analysis was specifically looking for decays through the known $^{14}$Be resonance unbound by 280 keV, that sets a lower limit to the 4-body decay energy. More details on the experimental set up and particle identification can be found in the original work from this experiment in Reference \cite{Snyder}.

\section{Analysis}

\begin{figure*}[htbp] 
   \includegraphics[width=7.0in]{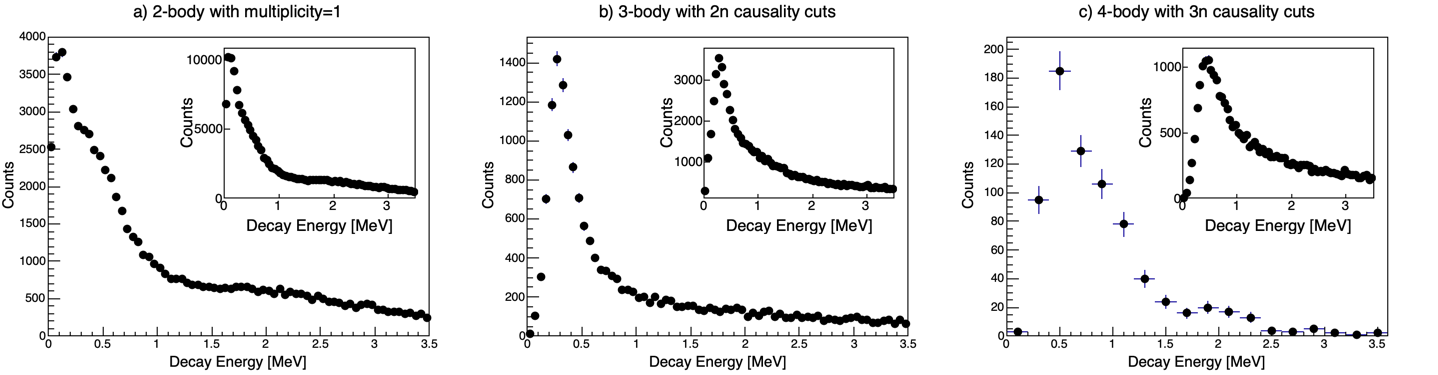} 
   \caption{Two-, three-, and four-body decay energies are presented from left to right, respectively. Panel a) shows the two-body decay energy with MoNA hit multiplicity required to be one. The inset is all two-body events. Panel b) is the three-body decay energy with 2n causality cuts applied. The inset is all three-body events. Panel c) is the four-body decay energy with the 3n causality cuts applied and with the three-body decay energy less than 1.0 MeV. The inset is all four-body events.}
   \label{fig:all}
\end{figure*}

Two-, three-, and four-body decay energy spectra were reconstructed as shown in Fig.~\ref{fig:all}. The two-body decay energy consisting of $^{12}$Be and one neutron shows three distinct features: a narrow peak around 110 keV, a structure around 400 keV, and a broad structure around 2 MeV. The 110 keV peak is significantly reduced relative to the other features when the hit multiplicity in MoNA is required to be exactly one. This spectrum looks similar to that seen in Figure 1 of Reference \cite{Kondo2010} and the authors attributed the peak to the inelastic excitation of the $^{14}$Be beam to its $2^+$ state which has been seen in \cite{Sugimoto2007, Aksyutina2013, Randisi2014}. The other features in this spectrum are previously observed states in $^{13}$Be \cite{Kondo2010, Randisi2014, Marks2015, Ribeiro2018, Corsi2019}. 

The three-body decay energy spectrum has a strong peak around 280 keV. However, decay energies with more than one neutron are subject to cross-talk events which can make single-neutron emissions appear as multi-neutron emissions. Applying causality conditions between the distance of the hits in MoNA and the velocity of these neutrons has been used to enhance the number of true three-body decays relative to cross talk events in several previous works from the MoNA Collaboration \cite{Hoffman2011,Spyrou2012, Lunderberg, Kohley13Li, Kohley10He, Smith12Be, Jones24O, Jones2015, Kuchera}. In this work, the minimum distance between the hits in MoNA was required to be 30 cm and the minimum relative velocity between the neutrons had to be greater than the average beam velocity of 10 cm/ns. When these causality conditions were applied, the high-energy shoulder on this peak was greatly diminished leaving a well-defined resonant shape consistent with the first excited state of $^{14}$Be \cite{Sugimoto2007}. The features identified as coming from $^{13}$Be in the two-body decay energy vanished when looking at events with multiplicity greater than one with the causality cuts applied. These conditions allowed the removal of $^{13}$Be events from the three- and four-body decay energy spectra. 

Previous attempts to reconstruct decay energies with more than two neutrons by the MoNA Collaboration have relied on simultaneous fitting of $n$-body decay energy spectra \cite{Kuchera, Jones2015, Sword}. The present analysis has sufficient statistics to apply the three-neutron (3n) causality conditions to analyze the four-body decay events. The 3n causality conditions are applied similarly to the 2n conditions but take into account all three of the combinations between neutrons one, two, and three. With the event selection applied to the four-body decay energy, a well-defined structure remained under 1 MeV. A measure of the effect of the causality cuts was determined from analyzing the simulated events. Without causality cuts, only $\approx 10\%$ of the events were events with three distinct neutrons. Once the 3n causality cuts were applied, the fraction of true 3n events was $\approx 60\%$. For the case of 2n emission, the causality cuts make the fraction of true 2n events $\approx 90\%$.

Monte Carlo simulations were performed to extract the properties of the observed resonances. Interactions of neutrons in MoNA were built in the Geant4 \cite{Geant4} framework with MENATE\_R \cite{Menate}. MENATE\_R models neutron interactions by including elastic and inelastic cross-sections of neutrons on protons or $^{12}$C, light output, and reproduces the resolution of position and time measurements of the MoNA bars. Each resonance is simulated with an $\ell$-dependent asymmetric Breit-Wigner line shape. We have used the same formalism defined in Equation 1 of Reference \cite{Jones2017}. However, due to the experimental energy resolution (see Fig.~\ref{fig:resolution}), there was no sensitivity to $\ell$ in the fitting. The reaction dynamics, decay processes, experimental acceptances, and resolutions of the detectors were modelled to provide a direct comparison between simulation and experiment. The four spectra involved in the simultaneous fitting were the four-body with 2n causality cuts, the four-body with 3n causality cuts, and the two- and three-body spectra with only the events that passed the conditions on the four-body spectra with 3n causality cuts. The unfiltered 4-body decay energy is shown in the inset of Fig.~\ref{fig:all}c. Figs. \ref{fig:14BeOnly} and \ref{fig:decayenergies} include the 4-body decay energy spectra with 2n and 3n causality cuts to see the effect that each cut has on the spectra and confirm the fits are consistent across the analysis. Additionally, the different cuts would enhance different features such as the presence of $^{14}$Be or $^{15}$Be in the spectrum. Only events with three-body decay energy less than 1.0 MeV were included to select events correlated to the $^{14}$Be* $2^+$ state. The free parameters in the fitting of the simulated states to the experimental data were the resonance energy and width for the $^{15}$Be state of interest and the relative scaling for each of the resonances. The simulations of $^{15}$Be to $^{12}$Be assumed a decay through $^{14}$Be*($2^+$) followed by a three-body phase space emission. This choice was guided by previous theoretical calculations \cite{Spyrou2011, Snyder, Fortune2015, Fortune2018} and experimental observation \cite{Kondo2010}. The direct population of $^{14}$Be*($2^+$) from inelastic excitation of the beam was also a contribution to the fit.


\section{Results}

\begin{figure*}[htbp] 
   \includegraphics[width=4.40in]{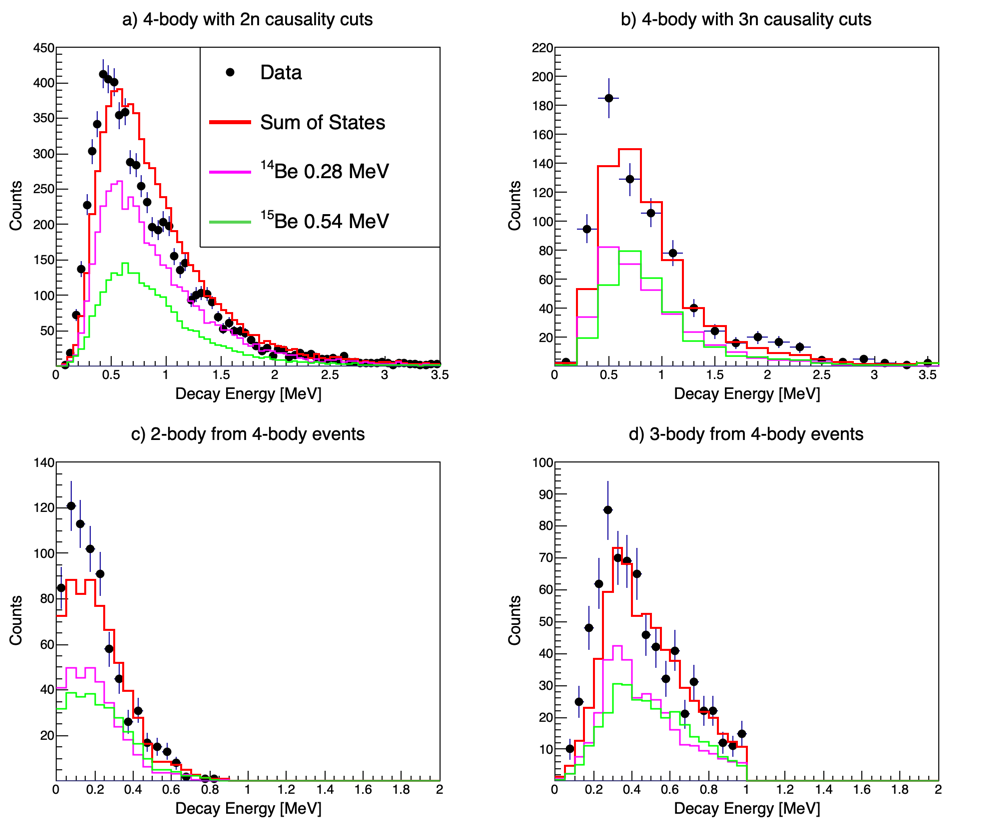} 
   \caption{The best fit only using the previously observed $^{14}$Be*($2^+$) (magenta) and $^{15}$Be (green) states. The data are shown by black markers and the sum of the simulated states is shown by the solid red line. The top panels are: a) the four-body decay energy with 2n causality cuts and b) four-body with 3n causality cuts and three-body decay energy less than 1.0 MeV. The bottom left and right panels are: c) the two-body and d) three-body decay energies, reconstructed from only the events that have made it through the 3n event selection.}
   \label{fig:14BeOnly}
\end{figure*}

\begin{figure*}[htbp] 
   \includegraphics[width=4.40in]{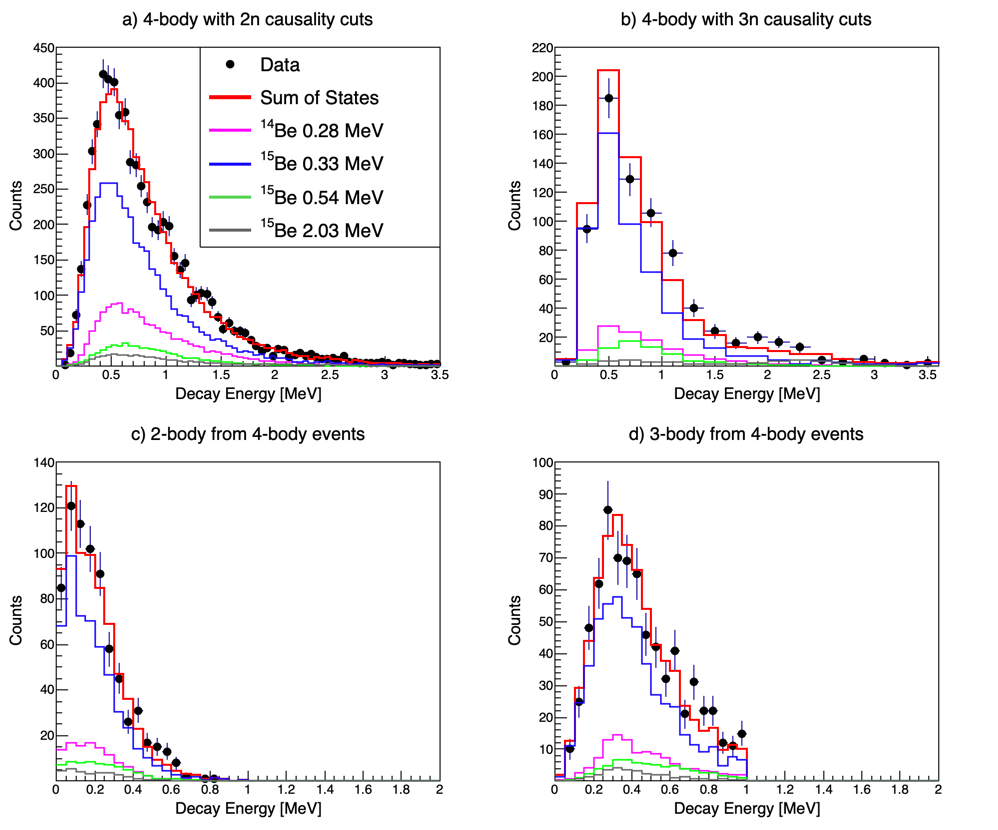} 
   \caption{The best fit with $^{15}$Be states (blue, green, gray) and the $^{14}$Be*($2^+$) state (magenta). The data are shown by black markers and the sum of the contributions from different states is shown by the solid red line. The top panels are: a) the four-body decay energy with 2n causality cuts and b) four-body with 3n causality cuts and three-body decay energy less than 1.0 MeV. The bottom left and right panels are: c) the two-body and d) three-body decay energies, reconstructed from only the events that have made it through the 3n event selection.}
   \label{fig:decayenergies}
\end{figure*}

When reconstructing events with $^{12}$Be in coincidence with neutrons, the three-body decay energy (shown in Fig.~\ref{fig:all}b) has a peak near the known first excited state of $^{14}$Be. Therefore, the first approach to fit the data only included the $^{14}$Be*($2^+$) state. The simulation of this state, which had a phase space decay with a three-body decay energy of 280 keV and width of 25 keV and $\ell=2$, alone failed to represent all four experimental spectra. The four-body decay energy with 2n causality cuts in the data was lower in energy than the simulation. The four-body decay energy strength with 3n causality cuts was significantly underrepresented by the simulation fitting to the data. The two-body decay energy strength was underrepresented and the three-body decay energy again had the peak of the data lower in energy.
 
The next approach kept the $^{14}$Be*($2^+$) state and included the previously observed $^{15}$Be resonance decaying through the $^{14}$Be*($2^+$) state with parameters ($E_{^{12}Be+3n}=540$ keV, $\Gamma=575$ keV \cite{Snyder}) shown in Fig.~\ref{fig:14BeOnly}. Panel a) shows $^{12}$Be events coincident with the first three time-ordered hits in MoNA that passed through the 2n causality conditions and had a three-body decay energy less than 1.0 MeV to reduce contributions of higher-energy states. Panel b) shows the same data with an additional set of causal conditions on the third hit in MoNA. The bottom two panels c) and d) are the two- and three-body decay energies reconstructed from the events with the 3n causality cuts. The key shows the color for each state included in the fit with the solid red line representing the sum of all contributions. The main peaks in the spectra again were not well represented by the simulations.

With the first two approaches unable to describe the data, simulations of new states in $^{15}$Be were performed with a range of energies and widths. The best fit determined by the minimum $\chi^2$ of the simulation and data is shown in Fig.~\ref{fig:decayenergies}. The spectra are dominated by a state in $^{15}$Be and the first excited state in $^{14}$Be that are 330 keV and 280 keV above the $^{12}$Be ground state, respectively. An additional $^{15}$Be state was included at 2 MeV which also decays through the $^{14}$Be*($2^+$) state to improve agreement with the high-energy data of the four-body decay spectra with causality cuts. Its strength was determined to be relatively weak compared to the low-energy state. The energy of this high-energy state is comparable to the high-energy structure included in the $^{14}$Be+n decay energy fit in Reference \cite{Snyder}.
 
A new state of $^{15}$Be unbound to $^{12}$Be by 330(20) keV, in addition to the two previously observed states and a higher-lying state, best described the data. An upper limit on the width of this state was determined by the minimization to be $\Gamma=200$ keV with an optimal value of $\Gamma=110$ keV. The relative strengths of the new state and the $^{14}$Be*($2^+$) state vary depending on the width. Comparable results are obtained when the strengths are nearly equal if the $^{15}$Be width is narrow ($< 10$ keV). The best fit with the width of 110 keV has the $^{15}$Be state 2.7 times stronger than the direct population of the $^{14}$Be*($2^+$) state (shown in Fig.~\ref{fig:decayenergies}). When the 3n causality cuts are applied to the simulated data for the $^{14}$Be state, the peak is significantly reduced compared to the data. This provides strong evidence for the need to include a $^{15}$Be resonance to describe the experimental data. The four decay energy spectra can only be simultaneously fitted when a low-energy state in $^{15}$Be is included. The spectra were not sensitive to the orbital angular momentum of the included resonance.

\begin{figure}[htbp] 
   \centering
   \includegraphics[width=7.6cm]{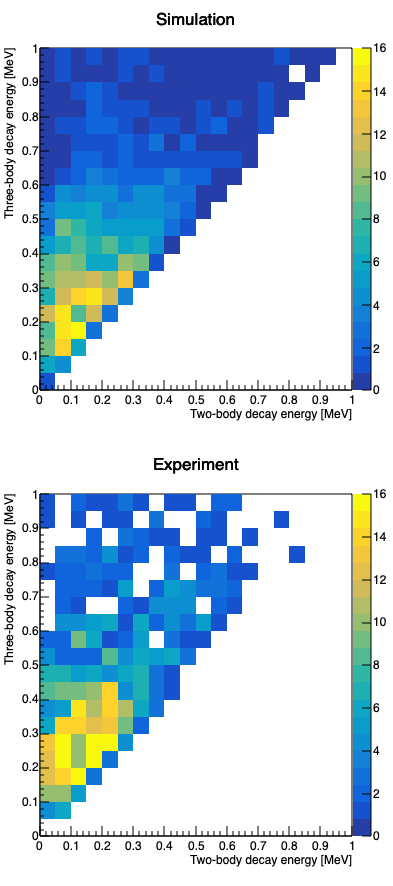} 
   \caption{Top panel: Simulated 3-body decay energy plotted on the vertical axis vs. the 2-body decay energy on the horizontal axis including 3n causality cuts. Bottom panel: Same parameters plotted for the experimental data.}
   \label{fig:3body2body}
\end{figure}

Fig.~\ref{fig:3body2body} shows a comparison between experimental data and simulation for correlations in two-dimensional space of three-body versus two-body decay energies on an event-by-event basis with causality cuts included. This is similar to what was done in the analysis for two-neutron emission from $^{24}$O \cite{Hoffman2011}. The bottom panel shows the experimental result while the top panel shows the result from the simulation which includes the $^{14}$Be and $^{15}$Be states discussed with the 1-D histogram fitting. This comparison provides additional evidence for the results obtained through decay energy fitting with detailed simulations.

\section{Discussion}
\begin{figure}[htbp] 
   \includegraphics[width=8.6cm]{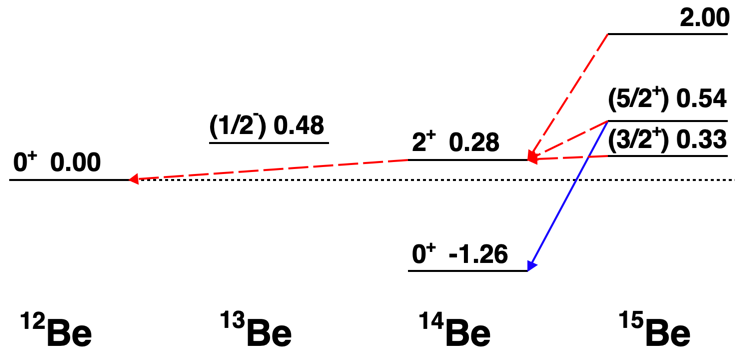} 
   \caption{The solid blue arrow represents the decay path of the previously observed $^{15}$Be state. The dashed red arrows represent the suggested decay paths used in this work. All energies listed are in MeV.}
   \label{fig:levels}
\end{figure}

Shell model calculations predicted $^{15}$Be to have a $3/2^+$ ground state and a nearby $5/2^+$ state \cite{Spyrou2011}. These calculations were done using the WBP Hamiltonian with restrictions that protons remained in the $p$-shell and that the neutron excitations were in the $p$- and $sd-$shells. The measurements of $^{15}$Be thus far have suggested the observed state to be the $5/2^+$ state \cite{Snyder, Belen}. This could mean that the newly observed state in this work is the $3/2^+$ ground state. It was also predicted that there is a strong overlap between the $3/2^+$ state and the first $2^+$ state in $^{14}$Be. While the spin-parity could not be confirmed in this work, the interpretation of a $3/2^+$ state sequentially decaying through the $2^+$ in $^{14}$Be followed by two-neutron emission to $^{12}$Be is consistent with the observations. It should also be pointed out that the best fit to data also includes a small branch of the previously observed $^{15}$Be state through this intermediate state in $^{14}$Be. This possibility was suggested in Reference \cite{Fortune2015}. A suggested decay scheme is shown in Fig.~\ref{fig:levels}.

One challenge in confirming that this new state is unique from the previously observed state is the uncertainty on the mass excess of $^{14}$Be. The 2020 Atomic Mass Evaluation reports a mass excess with an uncertainty of 130 keV for $^{14}$Be \cite{AME20}. The uncertainty on $^{12}$Be is only 1.9 keV \cite{AME20}. An improved mass measurement is needed to confirm that the resonance observed in this decay channel is unique from the previously observed resonance \cite{Snyder, Belen}.


\section{Conclusion}

Evidence for neutron unbound $^{15}$Be states has been presented from the reconstruction of events with $^{12}$Be and three neutrons. These events originated from a neutron-pick up reaction with a CD$_2$ target and $^{14}$Be beam. The energy of the dominant state, determined through simultaneous fitting of two-, three-, and four-body decay energy spectra simulations to data, is 330(20) keV. This is about 100 keV to 200 keV lower than the previously observed state decaying by one neutron emission to the $^{14}$Be ground state. \cite{Snyder, Belen}. Because of the energy of this state, it is a candidate for the ground state of $^{15}$Be. However, due to the uncertainty in the $^{14}$Be mass excess, improved mass measurements are needed to confirm this level ordering. If confirmed, these measurements would agree with shell model calculations and continue to rule out the possibility of a sequential decay of the ground state of $^{16}$Be, consistent with the observation of direct two-neutron emission \cite{Spyrou2012, Belen2024}.

\begin{acknowledgments}
The authors wish to thank the MoNA Collaboration and all who worked on the experiment to obtain the data. This work is supported by the National Science Foundation, USA under Grants No. PHY-2011398, No. PHY-2011265, No. PHY-1911418, and No. PHY-0606007.
\end{acknowledgments}

\bibliography{apssamp}

\begin{thebibliography}{35}%
\makeatletter
\providecommand \@ifxundefined [1]{%
 \@ifx{#1\undefined}
}%
\providecommand \@ifnum [1]{%
 \ifnum #1\expandafter \@firstoftwo
 \else \expandafter \@secondoftwo
 \fi
}%
\providecommand \@ifx [1]{%
 \ifx #1\expandafter \@firstoftwo
 \else \expandafter \@secondoftwo
 \fi
}%
\providecommand \natexlab [1]{#1}%
\providecommand \enquote  [1]{``#1''}%
\providecommand \bibnamefont  [1]{#1}%
\providecommand \bibfnamefont [1]{#1}%
\providecommand \citenamefont [1]{#1}%
\providecommand \href@noop [0]{\@secondoftwo}%
\providecommand \href [0]{\begingroup \@sanitize@url \@href}%
\providecommand \@href[1]{\@@startlink{#1}\@@href}%
\providecommand \@@href[1]{\endgroup#1\@@endlink}%
\providecommand \@sanitize@url [0]{\catcode `\\12\catcode `\$12\catcode
  `\&12\catcode `\#12\catcode `\^12\catcode `\_12\catcode `\%12\relax}%
\providecommand \@@startlink[1]{}%
\providecommand \@@endlink[0]{}%
\providecommand \url  [0]{\begingroup\@sanitize@url \@url }%
\providecommand \@url [1]{\endgroup\@href {#1}{\urlprefix }}%
\providecommand \urlprefix  [0]{URL }%
\providecommand \Eprint [0]{\href }%
\providecommand \doibase [0]{http://dx.doi.org/}%
\providecommand \selectlanguage [0]{\@gobble}%
\providecommand \bibinfo  [0]{\@secondoftwo}%
\providecommand \bibfield  [0]{\@secondoftwo}%
\providecommand \translation [1]{[#1]}%
\providecommand \BibitemOpen [0]{}%
\providecommand \bibitemStop [0]{}%
\providecommand \bibitemNoStop [0]{.\EOS\space}%
\providecommand \EOS [0]{\spacefactor3000\relax}%
\providecommand \BibitemShut  [1]{\csname bibitem#1\endcsname}%
\let\auto@bib@innerbib\@empty
\bibitem [{\citenamefont {Spyrou}\ \emph {et~al.}(2012)\citenamefont {Spyrou},
  \citenamefont {Kohley}, \citenamefont {Baumann}, \citenamefont {Bazin},
  \citenamefont {Brown}, \citenamefont {Christian}, \citenamefont {DeYoung},
  \citenamefont {Finck}, \citenamefont {Frank}, \citenamefont {Lunderberg},
  \citenamefont {Mosby}, \citenamefont {Peters}, \citenamefont {Schiller},
  \citenamefont {Smith}, \citenamefont {Snyder}, \citenamefont {Strongman},
  \citenamefont {Thoennessen},\ and\ \citenamefont {Volya}}]{Spyrou2012}%
  \BibitemOpen
  \bibfield  {author} {\bibinfo {author} {\bibfnamefont {A.}~\bibnamefont
  {Spyrou}}, \bibinfo {author} {\bibfnamefont {Z.}~\bibnamefont {Kohley}},
  \bibinfo {author} {\bibfnamefont {T.}~\bibnamefont {Baumann}}, \bibinfo
  {author} {\bibfnamefont {D.}~\bibnamefont {Bazin}}, \bibinfo {author}
  {\bibfnamefont {B.~A.}\ \bibnamefont {Brown}}, \bibinfo {author}
  {\bibfnamefont {G.}~\bibnamefont {Christian}}, \bibinfo {author}
  {\bibfnamefont {P.~A.}\ \bibnamefont {DeYoung}}, \bibinfo {author}
  {\bibfnamefont {J.~E.}\ \bibnamefont {Finck}}, \bibinfo {author}
  {\bibfnamefont {N.}~\bibnamefont {Frank}}, \bibinfo {author} {\bibfnamefont
  {E.}~\bibnamefont {Lunderberg}}, \bibinfo {author} {\bibfnamefont
  {S.}~\bibnamefont {Mosby}}, \bibinfo {author} {\bibfnamefont {W.~A.}\
  \bibnamefont {Peters}}, \bibinfo {author} {\bibfnamefont {A.}~\bibnamefont
  {Schiller}}, \bibinfo {author} {\bibfnamefont {J.~K.}\ \bibnamefont {Smith}},
  \bibinfo {author} {\bibfnamefont {J.}~\bibnamefont {Snyder}}, \bibinfo
  {author} {\bibfnamefont {M.~J.}\ \bibnamefont {Strongman}}, \bibinfo {author}
  {\bibfnamefont {M.}~\bibnamefont {Thoennessen}}, \ and\ \bibinfo {author}
  {\bibfnamefont {A.}~\bibnamefont {Volya}},\ }\href {\doibase
  10.1103/PhysRevLett.108.102501} {\bibfield  {journal} {\bibinfo  {journal}
  {Phys. Rev. Lett.}\ }\textbf {\bibinfo {volume} {108}},\ \bibinfo {pages}
  {102501} (\bibinfo {year} {2012})}\BibitemShut {NoStop}%
\bibitem [{\citenamefont {Marqu\'es}\ \emph {et~al.}(2012)\citenamefont
  {Marqu\'es}, \citenamefont {Orr}, \citenamefont {Achouri}, \citenamefont
  {Delaunay},\ and\ \citenamefont {Gibelin}}]{Marques2012}%
  \BibitemOpen
  \bibfield  {author} {\bibinfo {author} {\bibfnamefont {F.~M.}\ \bibnamefont
  {Marqu\'es}}, \bibinfo {author} {\bibfnamefont {N.~A.}\ \bibnamefont {Orr}},
  \bibinfo {author} {\bibfnamefont {N.~L.}\ \bibnamefont {Achouri}}, \bibinfo
  {author} {\bibfnamefont {F.}~\bibnamefont {Delaunay}}, \ and\ \bibinfo
  {author} {\bibfnamefont {J.}~\bibnamefont {Gibelin}},\ }\href {\doibase
  10.1103/PhysRevLett.109.239201} {\bibfield  {journal} {\bibinfo  {journal}
  {Phys. Rev. Lett.}\ }\textbf {\bibinfo {volume} {109}},\ \bibinfo {pages}
  {239201} (\bibinfo {year} {2012})}\BibitemShut {NoStop}%
\bibitem [{\citenamefont {Lovell}\ \emph {et~al.}(2017)\citenamefont {Lovell},
  \citenamefont {Nunes},\ and\ \citenamefont {Thompson}}]{Lovell2017}%
  \BibitemOpen
  \bibfield  {author} {\bibinfo {author} {\bibfnamefont {A.~E.}\ \bibnamefont
  {Lovell}}, \bibinfo {author} {\bibfnamefont {F.~M.}\ \bibnamefont {Nunes}}, \
  and\ \bibinfo {author} {\bibfnamefont {I.~J.}\ \bibnamefont {Thompson}},\
  }\href {\doibase 10.1103/PhysRevC.95.034605} {\bibfield  {journal} {\bibinfo
  {journal} {Phys. Rev. C}\ }\textbf {\bibinfo {volume} {95}},\ \bibinfo
  {pages} {034605} (\bibinfo {year} {2017})}\BibitemShut {NoStop}%
\bibitem [{\citenamefont {Casal}(2018)}]{Casal2018}%
  \BibitemOpen
  \bibfield  {author} {\bibinfo {author} {\bibfnamefont {J.}~\bibnamefont
  {Casal}},\ }\href {\doibase 10.1103/PhysRevC.97.034613} {\bibfield  {journal}
  {\bibinfo  {journal} {Phys. Rev. C}\ }\textbf {\bibinfo {volume} {97}},\
  \bibinfo {pages} {034613} (\bibinfo {year} {2018})}\BibitemShut {NoStop}%
\bibitem [{\citenamefont {Casal}\ and\ \citenamefont
  {G\'omez-Camacho}(2019)}]{Casal2019}%
  \BibitemOpen
  \bibfield  {author} {\bibinfo {author} {\bibfnamefont {J.}~\bibnamefont
  {Casal}}\ and\ \bibinfo {author} {\bibfnamefont {J.}~\bibnamefont
  {G\'omez-Camacho}},\ }\href {\doibase 10.1103/PhysRevC.99.014604} {\bibfield
  {journal} {\bibinfo  {journal} {Phys. Rev. C}\ }\textbf {\bibinfo {volume}
  {99}},\ \bibinfo {pages} {014604} (\bibinfo {year} {2019})}\BibitemShut
  {NoStop}%
\bibitem [{\citenamefont {Fortune}(2019)}]{Fortune2019}%
  \BibitemOpen
  \bibfield  {author} {\bibinfo {author} {\bibfnamefont {H.~T.}\ \bibnamefont
  {Fortune}},\ }\href {\doibase 10.1103/PhysRevC.99.044318} {\bibfield
  {journal} {\bibinfo  {journal} {Phys. Rev. C}\ }\textbf {\bibinfo {volume}
  {99}},\ \bibinfo {pages} {044318} (\bibinfo {year} {2019})}\BibitemShut
  {NoStop}%
\bibitem [{\citenamefont {Monteagudo}\ \emph {et~al.}(2024)\citenamefont
  {Monteagudo}, \citenamefont {Marqu\'es}, \citenamefont {Gibelin},
  \citenamefont {Orr}, \citenamefont {Corsi}, \citenamefont {Kubota},
  \citenamefont {Casal}, \citenamefont {G\'omez-Camacho}, \citenamefont
  {Authelet}, \citenamefont {Baba}, \citenamefont {Caesar}, \citenamefont
  {Calvet}, \citenamefont {Delbart}, \citenamefont {Dozono}, \citenamefont
  {Feng}, \citenamefont {Flavigny}, \citenamefont {Gheller}, \citenamefont
  {Giganon}, \citenamefont {Gillibert}, \citenamefont {Hasegawa}, \citenamefont
  {Isobe}, \citenamefont {Kanaya}, \citenamefont {Kawakami}, \citenamefont
  {Kim}, \citenamefont {Kiyokawa}, \citenamefont {Kobayashi}, \citenamefont
  {Kobayashi}, \citenamefont {Kobayashi}, \citenamefont {Kondo}, \citenamefont
  {Korkulu}, \citenamefont {Koyama}, \citenamefont {Lapoux}, \citenamefont
  {Maeda}, \citenamefont {Motobayashi}, \citenamefont {Miyazaki}, \citenamefont
  {Nakamura}, \citenamefont {Nakatsuka}, \citenamefont {Nishio}, \citenamefont
  {Obertelli}, \citenamefont {Ohkura}, \citenamefont {Ota}, \citenamefont
  {Otsu}, \citenamefont {Ozaki}, \citenamefont {Panin}, \citenamefont
  {Paschalis}, \citenamefont {Pollacco}, \citenamefont {Reichert},
  \citenamefont {Rousse}, \citenamefont {Saito}, \citenamefont {Sakaguchi},
  \citenamefont {Sako}, \citenamefont {Santamaria}, \citenamefont {Sasano},
  \citenamefont {Sato}, \citenamefont {Shikata}, \citenamefont {Shimizu},
  \citenamefont {Shindo}, \citenamefont {Stuhl}, \citenamefont {Sumikama},
  \citenamefont {Sun}, \citenamefont {Tabata}, \citenamefont {Togano},
  \citenamefont {Tsubota}, \citenamefont {Uesaka}, \citenamefont {Yang},
  \citenamefont {Yasuda}, \citenamefont {Yoneda},\ and\ \citenamefont
  {Zenihiro}}]{Belen2024}%
  \BibitemOpen
  \bibfield  {author} {\bibinfo {author} {\bibfnamefont {B.}~\bibnamefont
  {Monteagudo}}, \bibinfo {author} {\bibfnamefont {F.~M.}\ \bibnamefont
  {Marqu\'es}}, \bibinfo {author} {\bibfnamefont {J.}~\bibnamefont {Gibelin}},
  \bibinfo {author} {\bibfnamefont {N.~A.}\ \bibnamefont {Orr}}, \bibinfo
  {author} {\bibfnamefont {A.}~\bibnamefont {Corsi}}, \bibinfo {author}
  {\bibfnamefont {Y.}~\bibnamefont {Kubota}}, \bibinfo {author} {\bibfnamefont
  {J.}~\bibnamefont {Casal}}, \bibinfo {author} {\bibfnamefont
  {J.}~\bibnamefont {G\'omez-Camacho}}, \bibinfo {author} {\bibfnamefont
  {G.}~\bibnamefont {Authelet}}, \bibinfo {author} {\bibfnamefont
  {H.}~\bibnamefont {Baba}}, \bibinfo {author} {\bibfnamefont {C.}~\bibnamefont
  {Caesar}}, \bibinfo {author} {\bibfnamefont {D.}~\bibnamefont {Calvet}},
  \bibinfo {author} {\bibfnamefont {A.}~\bibnamefont {Delbart}}, \bibinfo
  {author} {\bibfnamefont {M.}~\bibnamefont {Dozono}}, \bibinfo {author}
  {\bibfnamefont {J.}~\bibnamefont {Feng}}, \bibinfo {author} {\bibfnamefont
  {F.}~\bibnamefont {Flavigny}}, \bibinfo {author} {\bibfnamefont {J.-M.}\
  \bibnamefont {Gheller}}, \bibinfo {author} {\bibfnamefont {A.}~\bibnamefont
  {Giganon}}, \bibinfo {author} {\bibfnamefont {A.}~\bibnamefont {Gillibert}},
  \bibinfo {author} {\bibfnamefont {K.}~\bibnamefont {Hasegawa}}, \bibinfo
  {author} {\bibfnamefont {T.}~\bibnamefont {Isobe}}, \bibinfo {author}
  {\bibfnamefont {Y.}~\bibnamefont {Kanaya}}, \bibinfo {author} {\bibfnamefont
  {S.}~\bibnamefont {Kawakami}}, \bibinfo {author} {\bibfnamefont
  {D.}~\bibnamefont {Kim}}, \bibinfo {author} {\bibfnamefont {Y.}~\bibnamefont
  {Kiyokawa}}, \bibinfo {author} {\bibfnamefont {M.}~\bibnamefont {Kobayashi}},
  \bibinfo {author} {\bibfnamefont {N.}~\bibnamefont {Kobayashi}}, \bibinfo
  {author} {\bibfnamefont {T.}~\bibnamefont {Kobayashi}}, \bibinfo {author}
  {\bibfnamefont {Y.}~\bibnamefont {Kondo}}, \bibinfo {author} {\bibfnamefont
  {Z.}~\bibnamefont {Korkulu}}, \bibinfo {author} {\bibfnamefont
  {S.}~\bibnamefont {Koyama}}, \bibinfo {author} {\bibfnamefont
  {V.}~\bibnamefont {Lapoux}}, \bibinfo {author} {\bibfnamefont
  {Y.}~\bibnamefont {Maeda}}, \bibinfo {author} {\bibfnamefont
  {T.}~\bibnamefont {Motobayashi}}, \bibinfo {author} {\bibfnamefont
  {T.}~\bibnamefont {Miyazaki}}, \bibinfo {author} {\bibfnamefont
  {T.}~\bibnamefont {Nakamura}}, \bibinfo {author} {\bibfnamefont
  {N.}~\bibnamefont {Nakatsuka}}, \bibinfo {author} {\bibfnamefont
  {Y.}~\bibnamefont {Nishio}}, \bibinfo {author} {\bibfnamefont
  {A.}~\bibnamefont {Obertelli}}, \bibinfo {author} {\bibfnamefont
  {A.}~\bibnamefont {Ohkura}}, \bibinfo {author} {\bibfnamefont
  {S.}~\bibnamefont {Ota}}, \bibinfo {author} {\bibfnamefont {H.}~\bibnamefont
  {Otsu}}, \bibinfo {author} {\bibfnamefont {T.}~\bibnamefont {Ozaki}},
  \bibinfo {author} {\bibfnamefont {V.}~\bibnamefont {Panin}}, \bibinfo
  {author} {\bibfnamefont {S.}~\bibnamefont {Paschalis}}, \bibinfo {author}
  {\bibfnamefont {E.~C.}\ \bibnamefont {Pollacco}}, \bibinfo {author}
  {\bibfnamefont {S.}~\bibnamefont {Reichert}}, \bibinfo {author}
  {\bibfnamefont {J.-Y.}\ \bibnamefont {Rousse}}, \bibinfo {author}
  {\bibfnamefont {A.~T.}\ \bibnamefont {Saito}}, \bibinfo {author}
  {\bibfnamefont {S.}~\bibnamefont {Sakaguchi}}, \bibinfo {author}
  {\bibfnamefont {M.}~\bibnamefont {Sako}}, \bibinfo {author} {\bibfnamefont
  {C.}~\bibnamefont {Santamaria}}, \bibinfo {author} {\bibfnamefont
  {M.}~\bibnamefont {Sasano}}, \bibinfo {author} {\bibfnamefont
  {H.}~\bibnamefont {Sato}}, \bibinfo {author} {\bibfnamefont {M.}~\bibnamefont
  {Shikata}}, \bibinfo {author} {\bibfnamefont {Y.}~\bibnamefont {Shimizu}},
  \bibinfo {author} {\bibfnamefont {Y.}~\bibnamefont {Shindo}}, \bibinfo
  {author} {\bibfnamefont {L.}~\bibnamefont {Stuhl}}, \bibinfo {author}
  {\bibfnamefont {T.}~\bibnamefont {Sumikama}}, \bibinfo {author}
  {\bibfnamefont {Y.~L.}\ \bibnamefont {Sun}}, \bibinfo {author} {\bibfnamefont
  {M.}~\bibnamefont {Tabata}}, \bibinfo {author} {\bibfnamefont
  {Y.}~\bibnamefont {Togano}}, \bibinfo {author} {\bibfnamefont
  {J.}~\bibnamefont {Tsubota}}, \bibinfo {author} {\bibfnamefont
  {T.}~\bibnamefont {Uesaka}}, \bibinfo {author} {\bibfnamefont {Z.~H.}\
  \bibnamefont {Yang}}, \bibinfo {author} {\bibfnamefont {J.}~\bibnamefont
  {Yasuda}}, \bibinfo {author} {\bibfnamefont {K.}~\bibnamefont {Yoneda}}, \
  and\ \bibinfo {author} {\bibfnamefont {J.}~\bibnamefont {Zenihiro}},\ }\href
  {\doibase 10.1103/PhysRevLett.132.082501} {\bibfield  {journal} {\bibinfo
  {journal} {Phys. Rev. Lett.}\ }\textbf {\bibinfo {volume} {132}},\ \bibinfo
  {pages} {082501} (\bibinfo {year} {2024})}\BibitemShut {NoStop}%
\bibitem [{\citenamefont {Spyrou}\ \emph {et~al.}(2011)\citenamefont {Spyrou},
  \citenamefont {Smith}, \citenamefont {Baumann}, \citenamefont {Brown},
  \citenamefont {Brown}, \citenamefont {Christian}, \citenamefont {DeYoung},
  \citenamefont {Frank}, \citenamefont {Mosby}, \citenamefont {Peters},
  \citenamefont {Schiller}, \citenamefont {Strongman}, \citenamefont
  {Thoennessen},\ and\ \citenamefont {Tostevin}}]{Spyrou2011}%
  \BibitemOpen
  \bibfield  {author} {\bibinfo {author} {\bibfnamefont {A.}~\bibnamefont
  {Spyrou}}, \bibinfo {author} {\bibfnamefont {J.~K.}\ \bibnamefont {Smith}},
  \bibinfo {author} {\bibfnamefont {T.}~\bibnamefont {Baumann}}, \bibinfo
  {author} {\bibfnamefont {B.~A.}\ \bibnamefont {Brown}}, \bibinfo {author}
  {\bibfnamefont {J.}~\bibnamefont {Brown}}, \bibinfo {author} {\bibfnamefont
  {G.}~\bibnamefont {Christian}}, \bibinfo {author} {\bibfnamefont {P.~A.}\
  \bibnamefont {DeYoung}}, \bibinfo {author} {\bibfnamefont {N.}~\bibnamefont
  {Frank}}, \bibinfo {author} {\bibfnamefont {S.}~\bibnamefont {Mosby}},
  \bibinfo {author} {\bibfnamefont {W.~A.}\ \bibnamefont {Peters}}, \bibinfo
  {author} {\bibfnamefont {A.}~\bibnamefont {Schiller}}, \bibinfo {author}
  {\bibfnamefont {M.~J.}\ \bibnamefont {Strongman}}, \bibinfo {author}
  {\bibfnamefont {M.}~\bibnamefont {Thoennessen}}, \ and\ \bibinfo {author}
  {\bibfnamefont {J.~A.}\ \bibnamefont {Tostevin}},\ }\href {\doibase
  10.1103/PhysRevC.84.044309} {\bibfield  {journal} {\bibinfo  {journal} {Phys.
  Rev. C}\ }\textbf {\bibinfo {volume} {84}},\ \bibinfo {pages} {044309}
  (\bibinfo {year} {2011})}\BibitemShut {NoStop}%
\bibitem [{\citenamefont {Snyder}\ \emph {et~al.}(2013)\citenamefont {Snyder},
  \citenamefont {Baumann}, \citenamefont {Christian}, \citenamefont
  {Haring-Kaye}, \citenamefont {DeYoung}, \citenamefont {Kohley}, \citenamefont
  {Luther}, \citenamefont {Mosby}, \citenamefont {Mosby}, \citenamefont
  {Simon}, \citenamefont {Smith}, \citenamefont {Spyrou}, \citenamefont
  {Stephenson},\ and\ \citenamefont {Thoennessen}}]{Snyder}%
  \BibitemOpen
  \bibfield  {author} {\bibinfo {author} {\bibfnamefont {J.}~\bibnamefont
  {Snyder}}, \bibinfo {author} {\bibfnamefont {T.}~\bibnamefont {Baumann}},
  \bibinfo {author} {\bibfnamefont {G.}~\bibnamefont {Christian}}, \bibinfo
  {author} {\bibfnamefont {R.~A.}\ \bibnamefont {Haring-Kaye}}, \bibinfo
  {author} {\bibfnamefont {P.~A.}\ \bibnamefont {DeYoung}}, \bibinfo {author}
  {\bibfnamefont {Z.}~\bibnamefont {Kohley}}, \bibinfo {author} {\bibfnamefont
  {B.}~\bibnamefont {Luther}}, \bibinfo {author} {\bibfnamefont
  {M.}~\bibnamefont {Mosby}}, \bibinfo {author} {\bibfnamefont
  {S.}~\bibnamefont {Mosby}}, \bibinfo {author} {\bibfnamefont
  {A.}~\bibnamefont {Simon}}, \bibinfo {author} {\bibfnamefont {J.~K.}\
  \bibnamefont {Smith}}, \bibinfo {author} {\bibfnamefont {A.}~\bibnamefont
  {Spyrou}}, \bibinfo {author} {\bibfnamefont {S.}~\bibnamefont {Stephenson}},
  \ and\ \bibinfo {author} {\bibfnamefont {M.}~\bibnamefont {Thoennessen}},\
  }\href {\doibase 10.1103/PhysRevC.88.031303} {\bibfield  {journal} {\bibinfo
  {journal} {Phys. Rev. C}\ }\textbf {\bibinfo {volume} {88}},\ \bibinfo
  {pages} {031303} (\bibinfo {year} {2013})}\BibitemShut {NoStop}%
\bibitem [{\citenamefont {Corsi}\ \emph {et~al.}(2021)\citenamefont {Corsi},
  \citenamefont {Monteagudo},\ and\ \citenamefont {Marqu{\'e}s}}]{Belen}%
  \BibitemOpen
  \bibfield  {author} {\bibinfo {author} {\bibfnamefont {A.}~\bibnamefont
  {Corsi}}, \bibinfo {author} {\bibfnamefont {B.}~\bibnamefont {Monteagudo}}, \
  and\ \bibinfo {author} {\bibfnamefont {F.~M.}\ \bibnamefont {Marqu{\'e}s}},\
  }\href {\doibase 10.1140/epja/s10050-021-00384-0} {\bibfield  {journal}
  {\bibinfo  {journal} {The European Physical Journal A}\ }\textbf {\bibinfo
  {volume} {57}},\ \bibinfo {pages} {88} (\bibinfo {year} {2021})}\BibitemShut
  {NoStop}%
\bibitem [{\citenamefont {Kuchera}\ \emph {et~al.}(2015)\citenamefont
  {Kuchera}, \citenamefont {Spyrou}, \citenamefont {Smith}, \citenamefont
  {Baumann}, \citenamefont {Christian}, \citenamefont {DeYoung}, \citenamefont
  {Finck}, \citenamefont {Frank}, \citenamefont {Jones}, \citenamefont
  {Kohley}, \citenamefont {Mosby}, \citenamefont {Peters},\ and\ \citenamefont
  {Thoennessen}}]{Kuchera}%
  \BibitemOpen
  \bibfield  {author} {\bibinfo {author} {\bibfnamefont {A.~N.}\ \bibnamefont
  {Kuchera}}, \bibinfo {author} {\bibfnamefont {A.}~\bibnamefont {Spyrou}},
  \bibinfo {author} {\bibfnamefont {J.~K.}\ \bibnamefont {Smith}}, \bibinfo
  {author} {\bibfnamefont {T.}~\bibnamefont {Baumann}}, \bibinfo {author}
  {\bibfnamefont {G.}~\bibnamefont {Christian}}, \bibinfo {author}
  {\bibfnamefont {P.~A.}\ \bibnamefont {DeYoung}}, \bibinfo {author}
  {\bibfnamefont {J.~E.}\ \bibnamefont {Finck}}, \bibinfo {author}
  {\bibfnamefont {N.}~\bibnamefont {Frank}}, \bibinfo {author} {\bibfnamefont
  {M.~D.}\ \bibnamefont {Jones}}, \bibinfo {author} {\bibfnamefont
  {Z.}~\bibnamefont {Kohley}}, \bibinfo {author} {\bibfnamefont
  {S.}~\bibnamefont {Mosby}}, \bibinfo {author} {\bibfnamefont {W.~A.}\
  \bibnamefont {Peters}}, \ and\ \bibinfo {author} {\bibfnamefont
  {M.}~\bibnamefont {Thoennessen}},\ }\href {\doibase
  10.1103/PhysRevC.91.017304} {\bibfield  {journal} {\bibinfo  {journal} {Phys.
  Rev. C}\ }\textbf {\bibinfo {volume} {91}},\ \bibinfo {pages} {017304}
  (\bibinfo {year} {2015})}\BibitemShut {NoStop}%
\bibitem [{\citenamefont {Fortune}(2015)}]{Fortune2015}%
  \BibitemOpen
  \bibfield  {author} {\bibinfo {author} {\bibfnamefont {H.~T.}\ \bibnamefont
  {Fortune}},\ }\href {\doibase 10.1103/PhysRevC.91.034314} {\bibfield
  {journal} {\bibinfo  {journal} {Phys. Rev. C}\ }\textbf {\bibinfo {volume}
  {91}},\ \bibinfo {pages} {034314} (\bibinfo {year} {2015})}\BibitemShut
  {NoStop}%
\bibitem [{\citenamefont {Fortune}(2018)}]{Fortune2018}%
  \BibitemOpen
  \bibfield  {author} {\bibinfo {author} {\bibfnamefont {H.~T.}\ \bibnamefont
  {Fortune}},\ }\href {\doibase 10.1103/PhysRevC.98.054317} {\bibfield
  {journal} {\bibinfo  {journal} {Phys. Rev. C}\ }\textbf {\bibinfo {volume}
  {98}},\ \bibinfo {pages} {054317} (\bibinfo {year} {2018})}\BibitemShut
  {NoStop}%
\bibitem [{\citenamefont {Bird}\ \emph {et~al.}(2005)\citenamefont {Bird},
  \citenamefont {Kenney}, \citenamefont {Toth}, \citenamefont {Weijers},
  \citenamefont {DeKamp}, \citenamefont {Thoennessen},\ and\ \citenamefont
  {Zeller}}]{sweeper}%
  \BibitemOpen
  \bibfield  {author} {\bibinfo {author} {\bibfnamefont {M.~D.}\ \bibnamefont
  {Bird}}, \bibinfo {author} {\bibfnamefont {S.~J.}\ \bibnamefont {Kenney}},
  \bibinfo {author} {\bibfnamefont {J.}~\bibnamefont {Toth}}, \bibinfo {author}
  {\bibfnamefont {H.~W.}\ \bibnamefont {Weijers}}, \bibinfo {author}
  {\bibfnamefont {J.~C.}\ \bibnamefont {DeKamp}}, \bibinfo {author}
  {\bibfnamefont {M.}~\bibnamefont {Thoennessen}}, \ and\ \bibinfo {author}
  {\bibfnamefont {A.~F.}\ \bibnamefont {Zeller}},\ }\href@noop {} {\bibfield
  {journal} {\bibinfo  {journal} {IEEE Transactions on Applied
  Superconductivity}\ }\textbf {\bibinfo {volume} {15}},\ \bibinfo {pages}
  {1252} (\bibinfo {year} {2005})}\BibitemShut {NoStop}%
\bibitem [{\citenamefont {Luther}\ \emph {et~al.}(2003)\citenamefont {Luther}
  \emph {et~al.}}]{mona1}%
  \BibitemOpen
  \bibfield  {author} {\bibinfo {author} {\bibfnamefont {B.}~\bibnamefont
  {Luther}} \emph {et~al.},\ }\href {\doibase
  http://dx.doi.org/10.1016/S0168-9002(03)01014-3} {\bibfield  {journal}
  {\bibinfo  {journal} {Nuclear Instruments and Methods in Physics Research
  Section A: Accelerators, Spectrometers, Detectors and Associated Equipment}\
  }\textbf {\bibinfo {volume} {505}},\ \bibinfo {pages} {33 } (\bibinfo {year}
  {2003})},\ \bibinfo {note} {{P}roceedings of the tenth Symposium on Radiation
  Measurements and Applications}\BibitemShut {NoStop}%
\bibitem [{\citenamefont {Baumann}\ \emph {et~al.}(2005)\citenamefont {Baumann}
  \emph {et~al.}}]{mona2}%
  \BibitemOpen
  \bibfield  {author} {\bibinfo {author} {\bibfnamefont {T.}~\bibnamefont
  {Baumann}} \emph {et~al.},\ }\href {\doibase
  http://dx.doi.org/10.1016/j.nima.2004.12.020} {\bibfield  {journal} {\bibinfo
   {journal} {Nuclear Instruments and Methods in Physics Research Section A:
  Accelerators, Spectrometers, Detectors and Associated Equipment}\ }\textbf
  {\bibinfo {volume} {543}},\ \bibinfo {pages} {517 } (\bibinfo {year}
  {2005})}\BibitemShut {NoStop}%
\bibitem [{\citenamefont {Kondo}\ \emph {et~al.}(2010)\citenamefont {Kondo},
  \citenamefont {Nakamura}, \citenamefont {Satou}, \citenamefont {Matsumoto},
  \citenamefont {Aoi}, \citenamefont {Endo}, \citenamefont {Fukuda},
  \citenamefont {Gomi}, \citenamefont {Hashimoto}, \citenamefont {Ishihara},
  \citenamefont {Kawai}, \citenamefont {Kitayama}, \citenamefont {Kobayashi},
  \citenamefont {Matsuda}, \citenamefont {Matsui}, \citenamefont {Motobayashi},
  \citenamefont {Nakabayashi}, \citenamefont {Okumura}, \citenamefont {Ong},
  \citenamefont {Onishi}, \citenamefont {Ogata}, \citenamefont {Otsu},
  \citenamefont {Sakurai}, \citenamefont {Shimoura}, \citenamefont {Shinohara},
  \citenamefont {Sugimoto}, \citenamefont {Takeuchi}, \citenamefont {Tamaki},
  \citenamefont {Togano},\ and\ \citenamefont {Yanagisawa}}]{Kondo2010}%
  \BibitemOpen
  \bibfield  {author} {\bibinfo {author} {\bibfnamefont {Y.}~\bibnamefont
  {Kondo}}, \bibinfo {author} {\bibfnamefont {T.}~\bibnamefont {Nakamura}},
  \bibinfo {author} {\bibfnamefont {Y.}~\bibnamefont {Satou}}, \bibinfo
  {author} {\bibfnamefont {T.}~\bibnamefont {Matsumoto}}, \bibinfo {author}
  {\bibfnamefont {N.}~\bibnamefont {Aoi}}, \bibinfo {author} {\bibfnamefont
  {N.}~\bibnamefont {Endo}}, \bibinfo {author} {\bibfnamefont {N.}~\bibnamefont
  {Fukuda}}, \bibinfo {author} {\bibfnamefont {T.}~\bibnamefont {Gomi}},
  \bibinfo {author} {\bibfnamefont {Y.}~\bibnamefont {Hashimoto}}, \bibinfo
  {author} {\bibfnamefont {M.}~\bibnamefont {Ishihara}}, \bibinfo {author}
  {\bibfnamefont {S.}~\bibnamefont {Kawai}}, \bibinfo {author} {\bibfnamefont
  {M.}~\bibnamefont {Kitayama}}, \bibinfo {author} {\bibfnamefont
  {T.}~\bibnamefont {Kobayashi}}, \bibinfo {author} {\bibfnamefont
  {Y.}~\bibnamefont {Matsuda}}, \bibinfo {author} {\bibfnamefont
  {N.}~\bibnamefont {Matsui}}, \bibinfo {author} {\bibfnamefont
  {T.}~\bibnamefont {Motobayashi}}, \bibinfo {author} {\bibfnamefont
  {T.}~\bibnamefont {Nakabayashi}}, \bibinfo {author} {\bibfnamefont
  {T.}~\bibnamefont {Okumura}}, \bibinfo {author} {\bibfnamefont
  {H.}~\bibnamefont {Ong}}, \bibinfo {author} {\bibfnamefont {T.}~\bibnamefont
  {Onishi}}, \bibinfo {author} {\bibfnamefont {K.}~\bibnamefont {Ogata}},
  \bibinfo {author} {\bibfnamefont {H.}~\bibnamefont {Otsu}}, \bibinfo {author}
  {\bibfnamefont {H.}~\bibnamefont {Sakurai}}, \bibinfo {author} {\bibfnamefont
  {S.}~\bibnamefont {Shimoura}}, \bibinfo {author} {\bibfnamefont
  {M.}~\bibnamefont {Shinohara}}, \bibinfo {author} {\bibfnamefont
  {T.}~\bibnamefont {Sugimoto}}, \bibinfo {author} {\bibfnamefont
  {S.}~\bibnamefont {Takeuchi}}, \bibinfo {author} {\bibfnamefont
  {M.}~\bibnamefont {Tamaki}}, \bibinfo {author} {\bibfnamefont
  {Y.}~\bibnamefont {Togano}}, \ and\ \bibinfo {author} {\bibfnamefont
  {Y.}~\bibnamefont {Yanagisawa}},\ }\href {\doibase
  https://doi.org/10.1016/j.physletb.2010.05.031} {\bibfield  {journal}
  {\bibinfo  {journal} {Physics Letters B}\ }\textbf {\bibinfo {volume}
  {690}},\ \bibinfo {pages} {245} (\bibinfo {year} {2010})}\BibitemShut
  {NoStop}%
\bibitem [{\citenamefont {Sugimoto}\ \emph {et~al.}(2007)\citenamefont
  {Sugimoto}, \citenamefont {Nakamura}, \citenamefont {Kondo}, \citenamefont
  {Aoi}, \citenamefont {Baba}, \citenamefont {Bazin}, \citenamefont {Fukuda},
  \citenamefont {Gomi}, \citenamefont {Hasegawa}, \citenamefont {Imai},
  \citenamefont {Ishihara}, \citenamefont {Kobayashi}, \citenamefont {Kubo},
  \citenamefont {Miura}, \citenamefont {Motobayashi}, \citenamefont {Otsu},
  \citenamefont {Saito}, \citenamefont {Sakurai}, \citenamefont {Shimoura},
  \citenamefont {Vinodkumar}, \citenamefont {Watanabe}, \citenamefont
  {Watanabe}, \citenamefont {Yakushiji}, \citenamefont {Yanagisawa},\ and\
  \citenamefont {Yoneda}}]{Sugimoto2007}%
  \BibitemOpen
  \bibfield  {author} {\bibinfo {author} {\bibfnamefont {T.}~\bibnamefont
  {Sugimoto}}, \bibinfo {author} {\bibfnamefont {T.}~\bibnamefont {Nakamura}},
  \bibinfo {author} {\bibfnamefont {Y.}~\bibnamefont {Kondo}}, \bibinfo
  {author} {\bibfnamefont {N.}~\bibnamefont {Aoi}}, \bibinfo {author}
  {\bibfnamefont {H.}~\bibnamefont {Baba}}, \bibinfo {author} {\bibfnamefont
  {D.}~\bibnamefont {Bazin}}, \bibinfo {author} {\bibfnamefont
  {N.}~\bibnamefont {Fukuda}}, \bibinfo {author} {\bibfnamefont
  {T.}~\bibnamefont {Gomi}}, \bibinfo {author} {\bibfnamefont {H.}~\bibnamefont
  {Hasegawa}}, \bibinfo {author} {\bibfnamefont {N.}~\bibnamefont {Imai}},
  \bibinfo {author} {\bibfnamefont {M.}~\bibnamefont {Ishihara}}, \bibinfo
  {author} {\bibfnamefont {T.}~\bibnamefont {Kobayashi}}, \bibinfo {author}
  {\bibfnamefont {T.}~\bibnamefont {Kubo}}, \bibinfo {author} {\bibfnamefont
  {M.}~\bibnamefont {Miura}}, \bibinfo {author} {\bibfnamefont
  {T.}~\bibnamefont {Motobayashi}}, \bibinfo {author} {\bibfnamefont
  {H.}~\bibnamefont {Otsu}}, \bibinfo {author} {\bibfnamefont {A.}~\bibnamefont
  {Saito}}, \bibinfo {author} {\bibfnamefont {H.}~\bibnamefont {Sakurai}},
  \bibinfo {author} {\bibfnamefont {S.}~\bibnamefont {Shimoura}}, \bibinfo
  {author} {\bibfnamefont {A.}~\bibnamefont {Vinodkumar}}, \bibinfo {author}
  {\bibfnamefont {K.}~\bibnamefont {Watanabe}}, \bibinfo {author}
  {\bibfnamefont {Y.}~\bibnamefont {Watanabe}}, \bibinfo {author}
  {\bibfnamefont {T.}~\bibnamefont {Yakushiji}}, \bibinfo {author}
  {\bibfnamefont {Y.}~\bibnamefont {Yanagisawa}}, \ and\ \bibinfo {author}
  {\bibfnamefont {K.}~\bibnamefont {Yoneda}},\ }\href {\doibase
  https://doi.org/10.1016/j.physletb.2007.08.052} {\bibfield  {journal}
  {\bibinfo  {journal} {Physics Letters B}\ }\textbf {\bibinfo {volume}
  {654}},\ \bibinfo {pages} {160} (\bibinfo {year} {2007})}\BibitemShut
  {NoStop}%
\bibitem [{\citenamefont {Aksyutina}\ \emph {et~al.}(2013)\citenamefont
  {Aksyutina}, \citenamefont {Aumann}, \citenamefont {Boretzky}, \citenamefont
  {Borge}, \citenamefont {Caesar}, \citenamefont {Chatillon}, \citenamefont
  {Chulkov}, \citenamefont {Cortina-Gil}, \citenamefont {Datta~Pramanik},
  \citenamefont {Emling}, \citenamefont {Fynbo}, \citenamefont {Geissel},
  \citenamefont {Heinz}, \citenamefont {Ickert}, \citenamefont {Johansson},
  \citenamefont {Jonson}, \citenamefont {Kulessa}, \citenamefont {Langer},
  \citenamefont {LeBleis}, \citenamefont {Mahata}, \citenamefont
  {M\"unzenberg}, \citenamefont {Nilsson}, \citenamefont {Nyman}, \citenamefont
  {Palit}, \citenamefont {Paschalis}, \citenamefont {Prokopowicz},
  \citenamefont {Reifarth}, \citenamefont {Rossi}, \citenamefont {Richter},
  \citenamefont {Riisager}, \citenamefont {Schrieder}, \citenamefont {Simon},
  \citenamefont {S\"ummerer}, \citenamefont {Tengblad}, \citenamefont {Thies},
  \citenamefont {Weick},\ and\ \citenamefont {Zhukov}}]{Aksyutina2013}%
  \BibitemOpen
  \bibfield  {author} {\bibinfo {author} {\bibfnamefont {Y.}~\bibnamefont
  {Aksyutina}}, \bibinfo {author} {\bibfnamefont {T.}~\bibnamefont {Aumann}},
  \bibinfo {author} {\bibfnamefont {K.}~\bibnamefont {Boretzky}}, \bibinfo
  {author} {\bibfnamefont {M.~J.~G.}\ \bibnamefont {Borge}}, \bibinfo {author}
  {\bibfnamefont {C.}~\bibnamefont {Caesar}}, \bibinfo {author} {\bibfnamefont
  {A.}~\bibnamefont {Chatillon}}, \bibinfo {author} {\bibfnamefont {L.~V.}\
  \bibnamefont {Chulkov}}, \bibinfo {author} {\bibfnamefont {D.}~\bibnamefont
  {Cortina-Gil}}, \bibinfo {author} {\bibfnamefont {U.}~\bibnamefont
  {Datta~Pramanik}}, \bibinfo {author} {\bibfnamefont {H.}~\bibnamefont
  {Emling}}, \bibinfo {author} {\bibfnamefont {H.~O.~U.}\ \bibnamefont
  {Fynbo}}, \bibinfo {author} {\bibfnamefont {H.}~\bibnamefont {Geissel}},
  \bibinfo {author} {\bibfnamefont {A.}~\bibnamefont {Heinz}}, \bibinfo
  {author} {\bibfnamefont {G.}~\bibnamefont {Ickert}}, \bibinfo {author}
  {\bibfnamefont {H.~T.}\ \bibnamefont {Johansson}}, \bibinfo {author}
  {\bibfnamefont {B.}~\bibnamefont {Jonson}}, \bibinfo {author} {\bibfnamefont
  {R.}~\bibnamefont {Kulessa}}, \bibinfo {author} {\bibfnamefont
  {C.}~\bibnamefont {Langer}}, \bibinfo {author} {\bibfnamefont
  {T.}~\bibnamefont {LeBleis}}, \bibinfo {author} {\bibfnamefont
  {K.}~\bibnamefont {Mahata}}, \bibinfo {author} {\bibfnamefont
  {G.}~\bibnamefont {M\"unzenberg}}, \bibinfo {author} {\bibfnamefont
  {T.}~\bibnamefont {Nilsson}}, \bibinfo {author} {\bibfnamefont
  {G.}~\bibnamefont {Nyman}}, \bibinfo {author} {\bibfnamefont
  {R.}~\bibnamefont {Palit}}, \bibinfo {author} {\bibfnamefont
  {S.}~\bibnamefont {Paschalis}}, \bibinfo {author} {\bibfnamefont
  {W.}~\bibnamefont {Prokopowicz}}, \bibinfo {author} {\bibfnamefont
  {R.}~\bibnamefont {Reifarth}}, \bibinfo {author} {\bibfnamefont
  {D.}~\bibnamefont {Rossi}}, \bibinfo {author} {\bibfnamefont
  {A.}~\bibnamefont {Richter}}, \bibinfo {author} {\bibfnamefont
  {K.}~\bibnamefont {Riisager}}, \bibinfo {author} {\bibfnamefont
  {G.}~\bibnamefont {Schrieder}}, \bibinfo {author} {\bibfnamefont
  {H.}~\bibnamefont {Simon}}, \bibinfo {author} {\bibfnamefont
  {K.}~\bibnamefont {S\"ummerer}}, \bibinfo {author} {\bibfnamefont
  {O.}~\bibnamefont {Tengblad}}, \bibinfo {author} {\bibfnamefont
  {R.}~\bibnamefont {Thies}}, \bibinfo {author} {\bibfnamefont
  {H.}~\bibnamefont {Weick}}, \ and\ \bibinfo {author} {\bibfnamefont {M.~V.}\
  \bibnamefont {Zhukov}},\ }\href {\doibase 10.1103/PhysRevLett.111.242501}
  {\bibfield  {journal} {\bibinfo  {journal} {Phys. Rev. Lett.}\ }\textbf
  {\bibinfo {volume} {111}},\ \bibinfo {pages} {242501} (\bibinfo {year}
  {2013})}\BibitemShut {NoStop}%
\bibitem [{\citenamefont {Randisi}\ \emph {et~al.}(2014)\citenamefont
  {Randisi}, \citenamefont {Leprince}, \citenamefont {Al~Falou}, \citenamefont
  {Orr}, \citenamefont {Marqu\'es}, \citenamefont {Achouri}, \citenamefont
  {Ang\'elique}, \citenamefont {Ashwood}, \citenamefont {Bastin}, \citenamefont
  {Bloxham}, \citenamefont {Brown}, \citenamefont {Catford}, \citenamefont
  {Curtis}, \citenamefont {Delaunay}, \citenamefont {Freer}, \citenamefont
  {de~G\'oes~Brennand}, \citenamefont {Haigh}, \citenamefont {Hanappe},
  \citenamefont {Harlin}, \citenamefont {Laurent}, \citenamefont {Lecouey},
  \citenamefont {Ninane}, \citenamefont {Patterson}, \citenamefont {Price},
  \citenamefont {Stuttg\'e},\ and\ \citenamefont {Thomas}}]{Randisi2014}%
  \BibitemOpen
  \bibfield  {author} {\bibinfo {author} {\bibfnamefont {G.}~\bibnamefont
  {Randisi}}, \bibinfo {author} {\bibfnamefont {A.}~\bibnamefont {Leprince}},
  \bibinfo {author} {\bibfnamefont {H.}~\bibnamefont {Al~Falou}}, \bibinfo
  {author} {\bibfnamefont {N.~A.}\ \bibnamefont {Orr}}, \bibinfo {author}
  {\bibfnamefont {F.~M.}\ \bibnamefont {Marqu\'es}}, \bibinfo {author}
  {\bibfnamefont {N.~L.}\ \bibnamefont {Achouri}}, \bibinfo {author}
  {\bibfnamefont {J.-C.}\ \bibnamefont {Ang\'elique}}, \bibinfo {author}
  {\bibfnamefont {N.}~\bibnamefont {Ashwood}}, \bibinfo {author} {\bibfnamefont
  {B.}~\bibnamefont {Bastin}}, \bibinfo {author} {\bibfnamefont
  {T.}~\bibnamefont {Bloxham}}, \bibinfo {author} {\bibfnamefont {B.~A.}\
  \bibnamefont {Brown}}, \bibinfo {author} {\bibfnamefont {W.~N.}\ \bibnamefont
  {Catford}}, \bibinfo {author} {\bibfnamefont {N.}~\bibnamefont {Curtis}},
  \bibinfo {author} {\bibfnamefont {F.}~\bibnamefont {Delaunay}}, \bibinfo
  {author} {\bibfnamefont {M.}~\bibnamefont {Freer}}, \bibinfo {author}
  {\bibfnamefont {E.}~\bibnamefont {de~G\'oes~Brennand}}, \bibinfo {author}
  {\bibfnamefont {P.}~\bibnamefont {Haigh}}, \bibinfo {author} {\bibfnamefont
  {F.}~\bibnamefont {Hanappe}}, \bibinfo {author} {\bibfnamefont
  {C.}~\bibnamefont {Harlin}}, \bibinfo {author} {\bibfnamefont
  {B.}~\bibnamefont {Laurent}}, \bibinfo {author} {\bibfnamefont {J.-L.}\
  \bibnamefont {Lecouey}}, \bibinfo {author} {\bibfnamefont {A.}~\bibnamefont
  {Ninane}}, \bibinfo {author} {\bibfnamefont {N.}~\bibnamefont {Patterson}},
  \bibinfo {author} {\bibfnamefont {D.}~\bibnamefont {Price}}, \bibinfo
  {author} {\bibfnamefont {L.}~\bibnamefont {Stuttg\'e}}, \ and\ \bibinfo
  {author} {\bibfnamefont {J.~S.}\ \bibnamefont {Thomas}},\ }\href {\doibase
  10.1103/PhysRevC.89.034320} {\bibfield  {journal} {\bibinfo  {journal} {Phys.
  Rev. C}\ }\textbf {\bibinfo {volume} {89}},\ \bibinfo {pages} {034320}
  (\bibinfo {year} {2014})}\BibitemShut {NoStop}%
\bibitem [{\citenamefont {Marks}\ \emph {et~al.}(2015)\citenamefont {Marks},
  \citenamefont {DeYoung}, \citenamefont {Smith}, \citenamefont {Baumann},
  \citenamefont {Brown}, \citenamefont {Frank}, \citenamefont {Hinnefeld},
  \citenamefont {Hoffman}, \citenamefont {Jones}, \citenamefont {Kohley},
  \citenamefont {Kuchera}, \citenamefont {Luther}, \citenamefont {Spyrou},
  \citenamefont {Stephenson}, \citenamefont {Sullivan}, \citenamefont
  {Thoennessen}, \citenamefont {Viscariello},\ and\ \citenamefont
  {Williams}}]{Marks2015}%
  \BibitemOpen
  \bibfield  {author} {\bibinfo {author} {\bibfnamefont {B.~R.}\ \bibnamefont
  {Marks}}, \bibinfo {author} {\bibfnamefont {P.~A.}\ \bibnamefont {DeYoung}},
  \bibinfo {author} {\bibfnamefont {J.~K.}\ \bibnamefont {Smith}}, \bibinfo
  {author} {\bibfnamefont {T.}~\bibnamefont {Baumann}}, \bibinfo {author}
  {\bibfnamefont {J.}~\bibnamefont {Brown}}, \bibinfo {author} {\bibfnamefont
  {N.}~\bibnamefont {Frank}}, \bibinfo {author} {\bibfnamefont
  {J.}~\bibnamefont {Hinnefeld}}, \bibinfo {author} {\bibfnamefont
  {M.}~\bibnamefont {Hoffman}}, \bibinfo {author} {\bibfnamefont {M.~D.}\
  \bibnamefont {Jones}}, \bibinfo {author} {\bibfnamefont {Z.}~\bibnamefont
  {Kohley}}, \bibinfo {author} {\bibfnamefont {A.~N.}\ \bibnamefont {Kuchera}},
  \bibinfo {author} {\bibfnamefont {B.}~\bibnamefont {Luther}}, \bibinfo
  {author} {\bibfnamefont {A.}~\bibnamefont {Spyrou}}, \bibinfo {author}
  {\bibfnamefont {S.}~\bibnamefont {Stephenson}}, \bibinfo {author}
  {\bibfnamefont {C.}~\bibnamefont {Sullivan}}, \bibinfo {author}
  {\bibfnamefont {M.}~\bibnamefont {Thoennessen}}, \bibinfo {author}
  {\bibfnamefont {N.}~\bibnamefont {Viscariello}}, \ and\ \bibinfo {author}
  {\bibfnamefont {S.~J.}\ \bibnamefont {Williams}},\ }\href {\doibase
  10.1103/PhysRevC.92.054320} {\bibfield  {journal} {\bibinfo  {journal} {Phys.
  Rev. C}\ }\textbf {\bibinfo {volume} {92}},\ \bibinfo {pages} {054320}
  (\bibinfo {year} {2015})}\BibitemShut {NoStop}%
\bibitem [{\citenamefont {Ribeiro}\ \emph {et~al.}(2018)\citenamefont
  {Ribeiro}, \citenamefont {N\'acher}, \citenamefont {Tengblad}, \citenamefont
  {D\'{\i}az~Fern\'andez}, \citenamefont {Aksyutina}, \citenamefont
  {Alvarez-Pol}, \citenamefont {Atar}, \citenamefont {Aumann}, \citenamefont
  {Avdeichikov}, \citenamefont {Beceiro-Novo}, \citenamefont {Bemmerer},
  \citenamefont {Benlliure}, \citenamefont {Bertulani}, \citenamefont
  {Boillos}, \citenamefont {Boretzky}, \citenamefont {Borge}, \citenamefont
  {Caamano}, \citenamefont {Caesar}, \citenamefont {Casarejos}, \citenamefont
  {Catford}, \citenamefont {Cederk\"all}, \citenamefont {Chartier},
  \citenamefont {Chulkov}, \citenamefont {Cortina-Gil}, \citenamefont {Cravo},
  \citenamefont {Crespo}, \citenamefont {Datta~Pramanik}, \citenamefont
  {Dillmann}, \citenamefont {Elekes}, \citenamefont {Enders}, \citenamefont
  {Ershova}, \citenamefont {Estrade}, \citenamefont {Farinon}, \citenamefont
  {Fraile}, \citenamefont {Freer}, \citenamefont {Fynbo}, \citenamefont
  {Galaviz}, \citenamefont {Geissel}, \citenamefont {Gernh\"auser},
  \citenamefont {Golubev}, \citenamefont {G\"obel}, \citenamefont {Hagdahl},
  \citenamefont {Heftrich}, \citenamefont {Heil}, \citenamefont {Heine},
  \citenamefont {Heinz}, \citenamefont {Henriques}, \citenamefont {Holl},
  \citenamefont {Hufnagel}, \citenamefont {Ignatov}, \citenamefont {Johansson},
  \citenamefont {Jonson}, \citenamefont {Kalantar-Nayestanaki}, \citenamefont
  {Kanungo}, \citenamefont {Kelic-Heil}, \citenamefont {Kurz}, \citenamefont
  {Kr\"oll}, \citenamefont {Labiche}, \citenamefont {Langer}, \citenamefont
  {Le~Bleis}, \citenamefont {Lemmon}, \citenamefont {Lindberg}, \citenamefont
  {Machado}, \citenamefont {Marganiec}, \citenamefont {Movsesyan},
  \citenamefont {Nilsson}, \citenamefont {Nociforo}, \citenamefont {Panin},
  \citenamefont {Paschalis}, \citenamefont {Perea}, \citenamefont {Petri},
  \citenamefont {Pietri}, \citenamefont {Plag}, \citenamefont {Reifarth},
  \citenamefont {Rigollet}, \citenamefont {Riisager}, \citenamefont {Rossi},
  \citenamefont {R\"oder}, \citenamefont {Savran}, \citenamefont {Scheit},
  \citenamefont {Simon}, \citenamefont {Sorlin}, \citenamefont {Syndikus},
  \citenamefont {Taylor}, \citenamefont {Thies}, \citenamefont {Velho},
  \citenamefont {Wagner}, \citenamefont {Wamers}, \citenamefont {Vandebrouck},
  \citenamefont {Weick}, \citenamefont {Wheldon}, \citenamefont {Wilson},
  \citenamefont {Wimmer}, \citenamefont {Winfield}, \citenamefont {Woods},
  \citenamefont {Zhukov}, \citenamefont {Zilges},\ and\ \citenamefont
  {Zuber}}]{Ribeiro2018}%
  \BibitemOpen
  \bibfield  {author} {\bibinfo {author} {\bibfnamefont {G.}~\bibnamefont
  {Ribeiro}}, \bibinfo {author} {\bibfnamefont {E.}~\bibnamefont {N\'acher}},
  \bibinfo {author} {\bibfnamefont {O.}~\bibnamefont {Tengblad}}, \bibinfo
  {author} {\bibfnamefont {P.}~\bibnamefont {D\'{\i}az~Fern\'andez}}, \bibinfo
  {author} {\bibfnamefont {Y.}~\bibnamefont {Aksyutina}}, \bibinfo {author}
  {\bibfnamefont {H.}~\bibnamefont {Alvarez-Pol}}, \bibinfo {author}
  {\bibfnamefont {L.}~\bibnamefont {Atar}}, \bibinfo {author} {\bibfnamefont
  {T.}~\bibnamefont {Aumann}}, \bibinfo {author} {\bibfnamefont
  {V.}~\bibnamefont {Avdeichikov}}, \bibinfo {author} {\bibfnamefont
  {S.}~\bibnamefont {Beceiro-Novo}}, \bibinfo {author} {\bibfnamefont
  {D.}~\bibnamefont {Bemmerer}}, \bibinfo {author} {\bibfnamefont
  {J.}~\bibnamefont {Benlliure}}, \bibinfo {author} {\bibfnamefont {C.~A.}\
  \bibnamefont {Bertulani}}, \bibinfo {author} {\bibfnamefont {J.~M.}\
  \bibnamefont {Boillos}}, \bibinfo {author} {\bibfnamefont {K.}~\bibnamefont
  {Boretzky}}, \bibinfo {author} {\bibfnamefont {M.~J.~G.}\ \bibnamefont
  {Borge}}, \bibinfo {author} {\bibfnamefont {M.}~\bibnamefont {Caamano}},
  \bibinfo {author} {\bibfnamefont {C.}~\bibnamefont {Caesar}}, \bibinfo
  {author} {\bibfnamefont {E.}~\bibnamefont {Casarejos}}, \bibinfo {author}
  {\bibfnamefont {W.}~\bibnamefont {Catford}}, \bibinfo {author} {\bibfnamefont
  {J.}~\bibnamefont {Cederk\"all}}, \bibinfo {author} {\bibfnamefont
  {M.}~\bibnamefont {Chartier}}, \bibinfo {author} {\bibfnamefont
  {L.}~\bibnamefont {Chulkov}}, \bibinfo {author} {\bibfnamefont
  {D.}~\bibnamefont {Cortina-Gil}}, \bibinfo {author} {\bibfnamefont
  {E.}~\bibnamefont {Cravo}}, \bibinfo {author} {\bibfnamefont
  {R.}~\bibnamefont {Crespo}}, \bibinfo {author} {\bibfnamefont
  {U.}~\bibnamefont {Datta~Pramanik}}, \bibinfo {author} {\bibfnamefont
  {I.}~\bibnamefont {Dillmann}}, \bibinfo {author} {\bibfnamefont
  {Z.}~\bibnamefont {Elekes}}, \bibinfo {author} {\bibfnamefont
  {J.}~\bibnamefont {Enders}}, \bibinfo {author} {\bibfnamefont
  {O.}~\bibnamefont {Ershova}}, \bibinfo {author} {\bibfnamefont
  {A.}~\bibnamefont {Estrade}}, \bibinfo {author} {\bibfnamefont
  {F.}~\bibnamefont {Farinon}}, \bibinfo {author} {\bibfnamefont {L.~M.}\
  \bibnamefont {Fraile}}, \bibinfo {author} {\bibfnamefont {M.}~\bibnamefont
  {Freer}}, \bibinfo {author} {\bibfnamefont {H.~O.~U.}\ \bibnamefont {Fynbo}},
  \bibinfo {author} {\bibfnamefont {D.}~\bibnamefont {Galaviz}}, \bibinfo
  {author} {\bibfnamefont {H.}~\bibnamefont {Geissel}}, \bibinfo {author}
  {\bibfnamefont {R.}~\bibnamefont {Gernh\"auser}}, \bibinfo {author}
  {\bibfnamefont {P.}~\bibnamefont {Golubev}}, \bibinfo {author} {\bibfnamefont
  {K.}~\bibnamefont {G\"obel}}, \bibinfo {author} {\bibfnamefont
  {J.}~\bibnamefont {Hagdahl}}, \bibinfo {author} {\bibfnamefont
  {T.}~\bibnamefont {Heftrich}}, \bibinfo {author} {\bibfnamefont
  {M.}~\bibnamefont {Heil}}, \bibinfo {author} {\bibfnamefont {M.}~\bibnamefont
  {Heine}}, \bibinfo {author} {\bibfnamefont {A.}~\bibnamefont {Heinz}},
  \bibinfo {author} {\bibfnamefont {A.}~\bibnamefont {Henriques}}, \bibinfo
  {author} {\bibfnamefont {M.}~\bibnamefont {Holl}}, \bibinfo {author}
  {\bibfnamefont {A.}~\bibnamefont {Hufnagel}}, \bibinfo {author}
  {\bibfnamefont {A.}~\bibnamefont {Ignatov}}, \bibinfo {author} {\bibfnamefont
  {H.~T.}\ \bibnamefont {Johansson}}, \bibinfo {author} {\bibfnamefont
  {B.}~\bibnamefont {Jonson}}, \bibinfo {author} {\bibfnamefont
  {N.}~\bibnamefont {Kalantar-Nayestanaki}}, \bibinfo {author} {\bibfnamefont
  {R.}~\bibnamefont {Kanungo}}, \bibinfo {author} {\bibfnamefont
  {A.}~\bibnamefont {Kelic-Heil}}, \bibinfo {author} {\bibfnamefont
  {N.}~\bibnamefont {Kurz}}, \bibinfo {author} {\bibfnamefont {T.}~\bibnamefont
  {Kr\"oll}}, \bibinfo {author} {\bibfnamefont {M.}~\bibnamefont {Labiche}},
  \bibinfo {author} {\bibfnamefont {C.}~\bibnamefont {Langer}}, \bibinfo
  {author} {\bibfnamefont {T.}~\bibnamefont {Le~Bleis}}, \bibinfo {author}
  {\bibfnamefont {R.}~\bibnamefont {Lemmon}}, \bibinfo {author} {\bibfnamefont
  {S.}~\bibnamefont {Lindberg}}, \bibinfo {author} {\bibfnamefont
  {J.}~\bibnamefont {Machado}}, \bibinfo {author} {\bibfnamefont
  {J.}~\bibnamefont {Marganiec}}, \bibinfo {author} {\bibfnamefont
  {A.}~\bibnamefont {Movsesyan}}, \bibinfo {author} {\bibfnamefont
  {T.}~\bibnamefont {Nilsson}}, \bibinfo {author} {\bibfnamefont
  {C.}~\bibnamefont {Nociforo}}, \bibinfo {author} {\bibfnamefont
  {V.}~\bibnamefont {Panin}}, \bibinfo {author} {\bibfnamefont
  {S.}~\bibnamefont {Paschalis}}, \bibinfo {author} {\bibfnamefont
  {A.}~\bibnamefont {Perea}}, \bibinfo {author} {\bibfnamefont
  {M.}~\bibnamefont {Petri}}, \bibinfo {author} {\bibfnamefont
  {S.}~\bibnamefont {Pietri}}, \bibinfo {author} {\bibfnamefont
  {R.}~\bibnamefont {Plag}}, \bibinfo {author} {\bibfnamefont {R.}~\bibnamefont
  {Reifarth}}, \bibinfo {author} {\bibfnamefont {C.}~\bibnamefont {Rigollet}},
  \bibinfo {author} {\bibfnamefont {K.}~\bibnamefont {Riisager}}, \bibinfo
  {author} {\bibfnamefont {D.}~\bibnamefont {Rossi}}, \bibinfo {author}
  {\bibfnamefont {M.}~\bibnamefont {R\"oder}}, \bibinfo {author} {\bibfnamefont
  {D.}~\bibnamefont {Savran}}, \bibinfo {author} {\bibfnamefont
  {H.}~\bibnamefont {Scheit}}, \bibinfo {author} {\bibfnamefont
  {H.}~\bibnamefont {Simon}}, \bibinfo {author} {\bibfnamefont
  {O.}~\bibnamefont {Sorlin}}, \bibinfo {author} {\bibfnamefont
  {I.}~\bibnamefont {Syndikus}}, \bibinfo {author} {\bibfnamefont {J.~T.}\
  \bibnamefont {Taylor}}, \bibinfo {author} {\bibfnamefont {R.}~\bibnamefont
  {Thies}}, \bibinfo {author} {\bibfnamefont {P.}~\bibnamefont {Velho}},
  \bibinfo {author} {\bibfnamefont {A.}~\bibnamefont {Wagner}}, \bibinfo
  {author} {\bibfnamefont {F.}~\bibnamefont {Wamers}}, \bibinfo {author}
  {\bibfnamefont {M.}~\bibnamefont {Vandebrouck}}, \bibinfo {author}
  {\bibfnamefont {H.}~\bibnamefont {Weick}}, \bibinfo {author} {\bibfnamefont
  {C.}~\bibnamefont {Wheldon}}, \bibinfo {author} {\bibfnamefont
  {G.}~\bibnamefont {Wilson}}, \bibinfo {author} {\bibfnamefont
  {C.}~\bibnamefont {Wimmer}}, \bibinfo {author} {\bibfnamefont {J.~S.}\
  \bibnamefont {Winfield}}, \bibinfo {author} {\bibfnamefont {P.}~\bibnamefont
  {Woods}}, \bibinfo {author} {\bibfnamefont {M.~V.}\ \bibnamefont {Zhukov}},
  \bibinfo {author} {\bibfnamefont {A.}~\bibnamefont {Zilges}}, \ and\ \bibinfo
  {author} {\bibfnamefont {K.}~\bibnamefont {Zuber}} (\bibinfo {collaboration}
  {${\mathrm{R}}^{3}\mathrm{B}$ Collaboration}),\ }\href {\doibase
  10.1103/PhysRevC.98.024603} {\bibfield  {journal} {\bibinfo  {journal} {Phys.
  Rev. C}\ }\textbf {\bibinfo {volume} {98}},\ \bibinfo {pages} {024603}
  (\bibinfo {year} {2018})}\BibitemShut {NoStop}%
\bibitem [{\citenamefont {Corsi}\ \emph {et~al.}(2019)\citenamefont {Corsi},
  \citenamefont {Kubota}, \citenamefont {Casal}, \citenamefont {Gómez-Ramos},
  \citenamefont {Moro}, \citenamefont {Authelet}, \citenamefont {Baba},
  \citenamefont {Caesar}, \citenamefont {Calvet}, \citenamefont {Delbart},
  \citenamefont {Dozono}, \citenamefont {Feng}, \citenamefont {Flavigny},
  \citenamefont {Gheller}, \citenamefont {Gibelin}, \citenamefont {Giganon},
  \citenamefont {Gillibert}, \citenamefont {Hasegawa}, \citenamefont {Isobe},
  \citenamefont {Kanaya}, \citenamefont {Kawakami}, \citenamefont {Kim},
  \citenamefont {Kiyokawa}, \citenamefont {Kobayashi}, \citenamefont
  {Kobayashi}, \citenamefont {Kobayashi}, \citenamefont {Kondo}, \citenamefont
  {Korkulu}, \citenamefont {Koyama}, \citenamefont {Lapoux}, \citenamefont
  {Maeda}, \citenamefont {Marqués}, \citenamefont {Motobayashi}, \citenamefont
  {Miyazaki}, \citenamefont {Nakamura}, \citenamefont {Nakatsuka},
  \citenamefont {Nishio}, \citenamefont {Obertelli}, \citenamefont {Ohkura},
  \citenamefont {Orr}, \citenamefont {Ota}, \citenamefont {Otsu}, \citenamefont
  {Ozaki}, \citenamefont {Panin}, \citenamefont {Paschalis}, \citenamefont
  {Pollacco}, \citenamefont {Reichert}, \citenamefont {Rousse}, \citenamefont
  {Saito}, \citenamefont {Sakaguchi}, \citenamefont {Sako}, \citenamefont
  {Santamaria}, \citenamefont {Sasano}, \citenamefont {Sato}, \citenamefont
  {Shikata}, \citenamefont {Shimizu}, \citenamefont {Shindo}, \citenamefont
  {Stuhl}, \citenamefont {Sumikama}, \citenamefont {Sun}, \citenamefont
  {Tabata}, \citenamefont {Togano}, \citenamefont {Tsubota}, \citenamefont
  {Uesaka}, \citenamefont {Yang}, \citenamefont {Yasuda}, \citenamefont
  {Yoneda},\ and\ \citenamefont {Zenihiro}}]{Corsi2019}%
  \BibitemOpen
  \bibfield  {author} {\bibinfo {author} {\bibfnamefont {A.}~\bibnamefont
  {Corsi}}, \bibinfo {author} {\bibfnamefont {Y.}~\bibnamefont {Kubota}},
  \bibinfo {author} {\bibfnamefont {J.}~\bibnamefont {Casal}}, \bibinfo
  {author} {\bibfnamefont {M.}~\bibnamefont {Gómez-Ramos}}, \bibinfo {author}
  {\bibfnamefont {A.}~\bibnamefont {Moro}}, \bibinfo {author} {\bibfnamefont
  {G.}~\bibnamefont {Authelet}}, \bibinfo {author} {\bibfnamefont
  {H.}~\bibnamefont {Baba}}, \bibinfo {author} {\bibfnamefont {C.}~\bibnamefont
  {Caesar}}, \bibinfo {author} {\bibfnamefont {D.}~\bibnamefont {Calvet}},
  \bibinfo {author} {\bibfnamefont {A.}~\bibnamefont {Delbart}}, \bibinfo
  {author} {\bibfnamefont {M.}~\bibnamefont {Dozono}}, \bibinfo {author}
  {\bibfnamefont {J.}~\bibnamefont {Feng}}, \bibinfo {author} {\bibfnamefont
  {F.}~\bibnamefont {Flavigny}}, \bibinfo {author} {\bibfnamefont {J.-M.}\
  \bibnamefont {Gheller}}, \bibinfo {author} {\bibfnamefont {J.}~\bibnamefont
  {Gibelin}}, \bibinfo {author} {\bibfnamefont {A.}~\bibnamefont {Giganon}},
  \bibinfo {author} {\bibfnamefont {A.}~\bibnamefont {Gillibert}}, \bibinfo
  {author} {\bibfnamefont {K.}~\bibnamefont {Hasegawa}}, \bibinfo {author}
  {\bibfnamefont {T.}~\bibnamefont {Isobe}}, \bibinfo {author} {\bibfnamefont
  {Y.}~\bibnamefont {Kanaya}}, \bibinfo {author} {\bibfnamefont
  {S.}~\bibnamefont {Kawakami}}, \bibinfo {author} {\bibfnamefont
  {D.}~\bibnamefont {Kim}}, \bibinfo {author} {\bibfnamefont {Y.}~\bibnamefont
  {Kiyokawa}}, \bibinfo {author} {\bibfnamefont {M.}~\bibnamefont {Kobayashi}},
  \bibinfo {author} {\bibfnamefont {N.}~\bibnamefont {Kobayashi}}, \bibinfo
  {author} {\bibfnamefont {T.}~\bibnamefont {Kobayashi}}, \bibinfo {author}
  {\bibfnamefont {Y.}~\bibnamefont {Kondo}}, \bibinfo {author} {\bibfnamefont
  {Z.}~\bibnamefont {Korkulu}}, \bibinfo {author} {\bibfnamefont
  {S.}~\bibnamefont {Koyama}}, \bibinfo {author} {\bibfnamefont
  {V.}~\bibnamefont {Lapoux}}, \bibinfo {author} {\bibfnamefont
  {Y.}~\bibnamefont {Maeda}}, \bibinfo {author} {\bibfnamefont
  {F.}~\bibnamefont {Marqués}}, \bibinfo {author} {\bibfnamefont
  {T.}~\bibnamefont {Motobayashi}}, \bibinfo {author} {\bibfnamefont
  {T.}~\bibnamefont {Miyazaki}}, \bibinfo {author} {\bibfnamefont
  {T.}~\bibnamefont {Nakamura}}, \bibinfo {author} {\bibfnamefont
  {N.}~\bibnamefont {Nakatsuka}}, \bibinfo {author} {\bibfnamefont
  {Y.}~\bibnamefont {Nishio}}, \bibinfo {author} {\bibfnamefont
  {A.}~\bibnamefont {Obertelli}}, \bibinfo {author} {\bibfnamefont
  {A.}~\bibnamefont {Ohkura}}, \bibinfo {author} {\bibfnamefont
  {N.}~\bibnamefont {Orr}}, \bibinfo {author} {\bibfnamefont {S.}~\bibnamefont
  {Ota}}, \bibinfo {author} {\bibfnamefont {H.}~\bibnamefont {Otsu}}, \bibinfo
  {author} {\bibfnamefont {T.}~\bibnamefont {Ozaki}}, \bibinfo {author}
  {\bibfnamefont {V.}~\bibnamefont {Panin}}, \bibinfo {author} {\bibfnamefont
  {S.}~\bibnamefont {Paschalis}}, \bibinfo {author} {\bibfnamefont
  {E.}~\bibnamefont {Pollacco}}, \bibinfo {author} {\bibfnamefont
  {S.}~\bibnamefont {Reichert}}, \bibinfo {author} {\bibfnamefont {J.-Y.}\
  \bibnamefont {Rousse}}, \bibinfo {author} {\bibfnamefont {A.}~\bibnamefont
  {Saito}}, \bibinfo {author} {\bibfnamefont {S.}~\bibnamefont {Sakaguchi}},
  \bibinfo {author} {\bibfnamefont {M.}~\bibnamefont {Sako}}, \bibinfo {author}
  {\bibfnamefont {C.}~\bibnamefont {Santamaria}}, \bibinfo {author}
  {\bibfnamefont {M.}~\bibnamefont {Sasano}}, \bibinfo {author} {\bibfnamefont
  {H.}~\bibnamefont {Sato}}, \bibinfo {author} {\bibfnamefont {M.}~\bibnamefont
  {Shikata}}, \bibinfo {author} {\bibfnamefont {Y.}~\bibnamefont {Shimizu}},
  \bibinfo {author} {\bibfnamefont {Y.}~\bibnamefont {Shindo}}, \bibinfo
  {author} {\bibfnamefont {L.}~\bibnamefont {Stuhl}}, \bibinfo {author}
  {\bibfnamefont {T.}~\bibnamefont {Sumikama}}, \bibinfo {author}
  {\bibfnamefont {Y.}~\bibnamefont {Sun}}, \bibinfo {author} {\bibfnamefont
  {M.}~\bibnamefont {Tabata}}, \bibinfo {author} {\bibfnamefont
  {Y.}~\bibnamefont {Togano}}, \bibinfo {author} {\bibfnamefont
  {J.}~\bibnamefont {Tsubota}}, \bibinfo {author} {\bibfnamefont
  {T.}~\bibnamefont {Uesaka}}, \bibinfo {author} {\bibfnamefont
  {Z.}~\bibnamefont {Yang}}, \bibinfo {author} {\bibfnamefont {J.}~\bibnamefont
  {Yasuda}}, \bibinfo {author} {\bibfnamefont {K.}~\bibnamefont {Yoneda}}, \
  and\ \bibinfo {author} {\bibfnamefont {J.}~\bibnamefont {Zenihiro}},\ }\href
  {\doibase https://doi.org/10.1016/j.physletb.2019.134843} {\bibfield
  {journal} {\bibinfo  {journal} {Physics Letters B}\ }\textbf {\bibinfo
  {volume} {797}},\ \bibinfo {pages} {134843} (\bibinfo {year}
  {2019})}\BibitemShut {NoStop}%
\bibitem [{\citenamefont {Hoffman}\ \emph {et~al.}(2011)\citenamefont
  {Hoffman}, \citenamefont {Baumann}, \citenamefont {Brown}, \citenamefont
  {DeYoung}, \citenamefont {Finck}, \citenamefont {Frank}, \citenamefont
  {Hinnefeld}, \citenamefont {Mosby}, \citenamefont {Peters}, \citenamefont
  {Rogers}, \citenamefont {Schiller}, \citenamefont {Snyder}, \citenamefont
  {Spyrou}, \citenamefont {Tabor},\ and\ \citenamefont
  {Thoennessen}}]{Hoffman2011}%
  \BibitemOpen
  \bibfield  {author} {\bibinfo {author} {\bibfnamefont {C.~R.}\ \bibnamefont
  {Hoffman}}, \bibinfo {author} {\bibfnamefont {T.}~\bibnamefont {Baumann}},
  \bibinfo {author} {\bibfnamefont {J.}~\bibnamefont {Brown}}, \bibinfo
  {author} {\bibfnamefont {P.~A.}\ \bibnamefont {DeYoung}}, \bibinfo {author}
  {\bibfnamefont {J.~E.}\ \bibnamefont {Finck}}, \bibinfo {author}
  {\bibfnamefont {N.}~\bibnamefont {Frank}}, \bibinfo {author} {\bibfnamefont
  {J.~D.}\ \bibnamefont {Hinnefeld}}, \bibinfo {author} {\bibfnamefont
  {S.}~\bibnamefont {Mosby}}, \bibinfo {author} {\bibfnamefont {W.~A.}\
  \bibnamefont {Peters}}, \bibinfo {author} {\bibfnamefont {W.~F.}\
  \bibnamefont {Rogers}}, \bibinfo {author} {\bibfnamefont {A.}~\bibnamefont
  {Schiller}}, \bibinfo {author} {\bibfnamefont {J.}~\bibnamefont {Snyder}},
  \bibinfo {author} {\bibfnamefont {A.}~\bibnamefont {Spyrou}}, \bibinfo
  {author} {\bibfnamefont {S.~L.}\ \bibnamefont {Tabor}}, \ and\ \bibinfo
  {author} {\bibfnamefont {M.}~\bibnamefont {Thoennessen}},\ }\href {\doibase
  10.1103/PhysRevC.83.031303} {\bibfield  {journal} {\bibinfo  {journal} {Phys.
  Rev. C}\ }\textbf {\bibinfo {volume} {83}},\ \bibinfo {pages} {031303}
  (\bibinfo {year} {2011})}\BibitemShut {NoStop}%
\bibitem [{\citenamefont {Lunderberg}\ \emph {et~al.}(2012)\citenamefont
  {Lunderberg}, \citenamefont {DeYoung}, \citenamefont {Kohley}, \citenamefont
  {Attanayake}, \citenamefont {Baumann}, \citenamefont {Bazin}, \citenamefont
  {Christian}, \citenamefont {Divaratne}, \citenamefont {Grimes}, \citenamefont
  {Haagsma}, \citenamefont {Finck}, \citenamefont {Frank}, \citenamefont
  {Luther}, \citenamefont {Mosby}, \citenamefont {Nagi}, \citenamefont
  {Peaslee}, \citenamefont {Schiller}, \citenamefont {Snyder}, \citenamefont
  {Spyrou}, \citenamefont {Strongman},\ and\ \citenamefont
  {Thoennessen}}]{Lunderberg}%
  \BibitemOpen
  \bibfield  {author} {\bibinfo {author} {\bibfnamefont {E.}~\bibnamefont
  {Lunderberg}}, \bibinfo {author} {\bibfnamefont {P.~A.}\ \bibnamefont
  {DeYoung}}, \bibinfo {author} {\bibfnamefont {Z.}~\bibnamefont {Kohley}},
  \bibinfo {author} {\bibfnamefont {H.}~\bibnamefont {Attanayake}}, \bibinfo
  {author} {\bibfnamefont {T.}~\bibnamefont {Baumann}}, \bibinfo {author}
  {\bibfnamefont {D.}~\bibnamefont {Bazin}}, \bibinfo {author} {\bibfnamefont
  {G.}~\bibnamefont {Christian}}, \bibinfo {author} {\bibfnamefont
  {D.}~\bibnamefont {Divaratne}}, \bibinfo {author} {\bibfnamefont {S.~M.}\
  \bibnamefont {Grimes}}, \bibinfo {author} {\bibfnamefont {A.}~\bibnamefont
  {Haagsma}}, \bibinfo {author} {\bibfnamefont {J.~E.}\ \bibnamefont {Finck}},
  \bibinfo {author} {\bibfnamefont {N.}~\bibnamefont {Frank}}, \bibinfo
  {author} {\bibfnamefont {B.}~\bibnamefont {Luther}}, \bibinfo {author}
  {\bibfnamefont {S.}~\bibnamefont {Mosby}}, \bibinfo {author} {\bibfnamefont
  {T.}~\bibnamefont {Nagi}}, \bibinfo {author} {\bibfnamefont {G.~F.}\
  \bibnamefont {Peaslee}}, \bibinfo {author} {\bibfnamefont {A.}~\bibnamefont
  {Schiller}}, \bibinfo {author} {\bibfnamefont {J.}~\bibnamefont {Snyder}},
  \bibinfo {author} {\bibfnamefont {A.}~\bibnamefont {Spyrou}}, \bibinfo
  {author} {\bibfnamefont {M.~J.}\ \bibnamefont {Strongman}}, \ and\ \bibinfo
  {author} {\bibfnamefont {M.}~\bibnamefont {Thoennessen}},\ }\href {\doibase
  10.1103/PhysRevLett.108.142503} {\bibfield  {journal} {\bibinfo  {journal}
  {Phys. Rev. Lett.}\ }\textbf {\bibinfo {volume} {108}},\ \bibinfo {pages}
  {142503} (\bibinfo {year} {2012})}\BibitemShut {NoStop}%
\bibitem [{\citenamefont {Kohley}\ \emph {et~al.}(2013)\citenamefont {Kohley},
  \citenamefont {Lunderberg}, \citenamefont {DeYoung}, \citenamefont {Volya},
  \citenamefont {Baumann}, \citenamefont {Bazin}, \citenamefont {Christian},
  \citenamefont {Cooper}, \citenamefont {Frank}, \citenamefont {Gade},
  \citenamefont {Hall}, \citenamefont {Hinnefeld}, \citenamefont {Luther},
  \citenamefont {Mosby}, \citenamefont {Peters}, \citenamefont {Smith},
  \citenamefont {Snyder}, \citenamefont {Spyrou},\ and\ \citenamefont
  {Thoennessen}}]{Kohley13Li}%
  \BibitemOpen
  \bibfield  {author} {\bibinfo {author} {\bibfnamefont {Z.}~\bibnamefont
  {Kohley}}, \bibinfo {author} {\bibfnamefont {E.}~\bibnamefont {Lunderberg}},
  \bibinfo {author} {\bibfnamefont {P.~A.}\ \bibnamefont {DeYoung}}, \bibinfo
  {author} {\bibfnamefont {A.}~\bibnamefont {Volya}}, \bibinfo {author}
  {\bibfnamefont {T.}~\bibnamefont {Baumann}}, \bibinfo {author} {\bibfnamefont
  {D.}~\bibnamefont {Bazin}}, \bibinfo {author} {\bibfnamefont
  {G.}~\bibnamefont {Christian}}, \bibinfo {author} {\bibfnamefont {N.~L.}\
  \bibnamefont {Cooper}}, \bibinfo {author} {\bibfnamefont {N.}~\bibnamefont
  {Frank}}, \bibinfo {author} {\bibfnamefont {A.}~\bibnamefont {Gade}},
  \bibinfo {author} {\bibfnamefont {C.}~\bibnamefont {Hall}}, \bibinfo {author}
  {\bibfnamefont {J.}~\bibnamefont {Hinnefeld}}, \bibinfo {author}
  {\bibfnamefont {B.}~\bibnamefont {Luther}}, \bibinfo {author} {\bibfnamefont
  {S.}~\bibnamefont {Mosby}}, \bibinfo {author} {\bibfnamefont {W.~A.}\
  \bibnamefont {Peters}}, \bibinfo {author} {\bibfnamefont {J.~K.}\
  \bibnamefont {Smith}}, \bibinfo {author} {\bibfnamefont {J.}~\bibnamefont
  {Snyder}}, \bibinfo {author} {\bibfnamefont {A.}~\bibnamefont {Spyrou}}, \
  and\ \bibinfo {author} {\bibfnamefont {M.}~\bibnamefont {Thoennessen}},\
  }\href {\doibase 10.1103/PhysRevC.87.011304} {\bibfield  {journal} {\bibinfo
  {journal} {Phys. Rev. C}\ }\textbf {\bibinfo {volume} {87}},\ \bibinfo
  {pages} {011304} (\bibinfo {year} {2013})}\BibitemShut {NoStop}%
\bibitem [{\citenamefont {Kohley}\ \emph {et~al.}(2012)\citenamefont {Kohley},
  \citenamefont {Snyder}, \citenamefont {Baumann}, \citenamefont {Christian},
  \citenamefont {DeYoung}, \citenamefont {Finck}, \citenamefont {Haring-Kaye},
  \citenamefont {Jones}, \citenamefont {Lunderberg}, \citenamefont {Luther},
  \citenamefont {Mosby}, \citenamefont {Simon}, \citenamefont {Smith},
  \citenamefont {Spyrou}, \citenamefont {Stephenson},\ and\ \citenamefont
  {Thoennessen}}]{Kohley10He}%
  \BibitemOpen
  \bibfield  {author} {\bibinfo {author} {\bibfnamefont {Z.}~\bibnamefont
  {Kohley}}, \bibinfo {author} {\bibfnamefont {J.}~\bibnamefont {Snyder}},
  \bibinfo {author} {\bibfnamefont {T.}~\bibnamefont {Baumann}}, \bibinfo
  {author} {\bibfnamefont {G.}~\bibnamefont {Christian}}, \bibinfo {author}
  {\bibfnamefont {P.~A.}\ \bibnamefont {DeYoung}}, \bibinfo {author}
  {\bibfnamefont {J.~E.}\ \bibnamefont {Finck}}, \bibinfo {author}
  {\bibfnamefont {R.~A.}\ \bibnamefont {Haring-Kaye}}, \bibinfo {author}
  {\bibfnamefont {M.}~\bibnamefont {Jones}}, \bibinfo {author} {\bibfnamefont
  {E.}~\bibnamefont {Lunderberg}}, \bibinfo {author} {\bibfnamefont
  {B.}~\bibnamefont {Luther}}, \bibinfo {author} {\bibfnamefont
  {S.}~\bibnamefont {Mosby}}, \bibinfo {author} {\bibfnamefont
  {A.}~\bibnamefont {Simon}}, \bibinfo {author} {\bibfnamefont {J.~K.}\
  \bibnamefont {Smith}}, \bibinfo {author} {\bibfnamefont {A.}~\bibnamefont
  {Spyrou}}, \bibinfo {author} {\bibfnamefont {S.~L.}\ \bibnamefont
  {Stephenson}}, \ and\ \bibinfo {author} {\bibfnamefont {M.}~\bibnamefont
  {Thoennessen}},\ }\href {\doibase 10.1103/PhysRevLett.109.232501} {\bibfield
  {journal} {\bibinfo  {journal} {Phys. Rev. Lett.}\ }\textbf {\bibinfo
  {volume} {109}},\ \bibinfo {pages} {232501} (\bibinfo {year}
  {2012})}\BibitemShut {NoStop}%
\bibitem [{\citenamefont {Smith}\ \emph {et~al.}(2014)\citenamefont {Smith},
  \citenamefont {Baumann}, \citenamefont {Bazin}, \citenamefont {Brown},
  \citenamefont {Casarotto}, \citenamefont {DeYoung}, \citenamefont {Frank},
  \citenamefont {Hinnefeld}, \citenamefont {Hoffman}, \citenamefont {Jones},
  \citenamefont {Kohley}, \citenamefont {Luther}, \citenamefont {Marks},
  \citenamefont {Smith}, \citenamefont {Snyder}, \citenamefont {Spyrou},
  \citenamefont {Stephenson}, \citenamefont {Thoennessen}, \citenamefont
  {Viscariello},\ and\ \citenamefont {Williams}}]{Smith12Be}%
  \BibitemOpen
  \bibfield  {author} {\bibinfo {author} {\bibfnamefont {J.~K.}\ \bibnamefont
  {Smith}}, \bibinfo {author} {\bibfnamefont {T.}~\bibnamefont {Baumann}},
  \bibinfo {author} {\bibfnamefont {D.}~\bibnamefont {Bazin}}, \bibinfo
  {author} {\bibfnamefont {J.}~\bibnamefont {Brown}}, \bibinfo {author}
  {\bibfnamefont {S.}~\bibnamefont {Casarotto}}, \bibinfo {author}
  {\bibfnamefont {P.~A.}\ \bibnamefont {DeYoung}}, \bibinfo {author}
  {\bibfnamefont {N.}~\bibnamefont {Frank}}, \bibinfo {author} {\bibfnamefont
  {J.}~\bibnamefont {Hinnefeld}}, \bibinfo {author} {\bibfnamefont
  {M.}~\bibnamefont {Hoffman}}, \bibinfo {author} {\bibfnamefont {M.~D.}\
  \bibnamefont {Jones}}, \bibinfo {author} {\bibfnamefont {Z.}~\bibnamefont
  {Kohley}}, \bibinfo {author} {\bibfnamefont {B.}~\bibnamefont {Luther}},
  \bibinfo {author} {\bibfnamefont {B.}~\bibnamefont {Marks}}, \bibinfo
  {author} {\bibfnamefont {N.}~\bibnamefont {Smith}}, \bibinfo {author}
  {\bibfnamefont {J.}~\bibnamefont {Snyder}}, \bibinfo {author} {\bibfnamefont
  {A.}~\bibnamefont {Spyrou}}, \bibinfo {author} {\bibfnamefont {S.~L.}\
  \bibnamefont {Stephenson}}, \bibinfo {author} {\bibfnamefont
  {M.}~\bibnamefont {Thoennessen}}, \bibinfo {author} {\bibfnamefont
  {N.}~\bibnamefont {Viscariello}}, \ and\ \bibinfo {author} {\bibfnamefont
  {S.~J.}\ \bibnamefont {Williams}},\ }\href {\doibase
  10.1103/PhysRevC.90.024309} {\bibfield  {journal} {\bibinfo  {journal} {Phys.
  Rev. C}\ }\textbf {\bibinfo {volume} {90}},\ \bibinfo {pages} {024309}
  (\bibinfo {year} {2014})}\BibitemShut {NoStop}%
\bibitem [{\citenamefont {Jones}\ \emph
  {et~al.}(2015{\natexlab{a}})\citenamefont {Jones}, \citenamefont {Frank},
  \citenamefont {Baumann}, \citenamefont {Brett}, \citenamefont {Bullaro},
  \citenamefont {DeYoung}, \citenamefont {Finck}, \citenamefont {Hammerton},
  \citenamefont {Hinnefeld}, \citenamefont {Kohley}, \citenamefont {Kuchera},
  \citenamefont {Pereira}, \citenamefont {Rabeh}, \citenamefont {Rogers},
  \citenamefont {Smith}, \citenamefont {Spyrou}, \citenamefont {Stephenson},
  \citenamefont {Stiefel}, \citenamefont {Tuttle-Timm}, \citenamefont
  {Zegers},\ and\ \citenamefont {Thoennessen}}]{Jones24O}%
  \BibitemOpen
  \bibfield  {author} {\bibinfo {author} {\bibfnamefont {M.~D.}\ \bibnamefont
  {Jones}}, \bibinfo {author} {\bibfnamefont {N.}~\bibnamefont {Frank}},
  \bibinfo {author} {\bibfnamefont {T.}~\bibnamefont {Baumann}}, \bibinfo
  {author} {\bibfnamefont {J.}~\bibnamefont {Brett}}, \bibinfo {author}
  {\bibfnamefont {J.}~\bibnamefont {Bullaro}}, \bibinfo {author} {\bibfnamefont
  {P.~A.}\ \bibnamefont {DeYoung}}, \bibinfo {author} {\bibfnamefont {J.~E.}\
  \bibnamefont {Finck}}, \bibinfo {author} {\bibfnamefont {K.}~\bibnamefont
  {Hammerton}}, \bibinfo {author} {\bibfnamefont {J.}~\bibnamefont
  {Hinnefeld}}, \bibinfo {author} {\bibfnamefont {Z.}~\bibnamefont {Kohley}},
  \bibinfo {author} {\bibfnamefont {A.~N.}\ \bibnamefont {Kuchera}}, \bibinfo
  {author} {\bibfnamefont {J.}~\bibnamefont {Pereira}}, \bibinfo {author}
  {\bibfnamefont {A.}~\bibnamefont {Rabeh}}, \bibinfo {author} {\bibfnamefont
  {W.~F.}\ \bibnamefont {Rogers}}, \bibinfo {author} {\bibfnamefont {J.~K.}\
  \bibnamefont {Smith}}, \bibinfo {author} {\bibfnamefont {A.}~\bibnamefont
  {Spyrou}}, \bibinfo {author} {\bibfnamefont {S.~L.}\ \bibnamefont
  {Stephenson}}, \bibinfo {author} {\bibfnamefont {K.}~\bibnamefont {Stiefel}},
  \bibinfo {author} {\bibfnamefont {M.}~\bibnamefont {Tuttle-Timm}}, \bibinfo
  {author} {\bibfnamefont {R.~G.~T.}\ \bibnamefont {Zegers}}, \ and\ \bibinfo
  {author} {\bibfnamefont {M.}~\bibnamefont {Thoennessen}},\ }\href {\doibase
  10.1103/PhysRevC.92.051306} {\bibfield  {journal} {\bibinfo  {journal} {Phys.
  Rev. C}\ }\textbf {\bibinfo {volume} {92}},\ \bibinfo {pages} {051306}
  (\bibinfo {year} {2015}{\natexlab{a}})}\BibitemShut {NoStop}%
\bibitem [{\citenamefont {Jones}\ \emph
  {et~al.}(2015{\natexlab{b}})\citenamefont {Jones}, \citenamefont {Kohley},
  \citenamefont {Baumann}, \citenamefont {Christian}, \citenamefont {DeYoung},
  \citenamefont {Finck}, \citenamefont {Frank}, \citenamefont {Haring-Kaye},
  \citenamefont {Kuchera}, \citenamefont {Luther}, \citenamefont {Mosby},
  \citenamefont {Smith}, \citenamefont {Snyder}, \citenamefont {Spyrou},
  \citenamefont {Stephenson},\ and\ \citenamefont {Thoennessen}}]{Jones2015}%
  \BibitemOpen
  \bibfield  {author} {\bibinfo {author} {\bibfnamefont {M.~D.}\ \bibnamefont
  {Jones}}, \bibinfo {author} {\bibfnamefont {Z.}~\bibnamefont {Kohley}},
  \bibinfo {author} {\bibfnamefont {T.}~\bibnamefont {Baumann}}, \bibinfo
  {author} {\bibfnamefont {G.}~\bibnamefont {Christian}}, \bibinfo {author}
  {\bibfnamefont {P.~A.}\ \bibnamefont {DeYoung}}, \bibinfo {author}
  {\bibfnamefont {J.~E.}\ \bibnamefont {Finck}}, \bibinfo {author}
  {\bibfnamefont {N.}~\bibnamefont {Frank}}, \bibinfo {author} {\bibfnamefont
  {R.~A.}\ \bibnamefont {Haring-Kaye}}, \bibinfo {author} {\bibfnamefont
  {A.~N.}\ \bibnamefont {Kuchera}}, \bibinfo {author} {\bibfnamefont
  {B.}~\bibnamefont {Luther}}, \bibinfo {author} {\bibfnamefont
  {S.}~\bibnamefont {Mosby}}, \bibinfo {author} {\bibfnamefont {J.~K.}\
  \bibnamefont {Smith}}, \bibinfo {author} {\bibfnamefont {J.}~\bibnamefont
  {Snyder}}, \bibinfo {author} {\bibfnamefont {A.}~\bibnamefont {Spyrou}},
  \bibinfo {author} {\bibfnamefont {S.~L.}\ \bibnamefont {Stephenson}}, \ and\
  \bibinfo {author} {\bibfnamefont {M.}~\bibnamefont {Thoennessen}},\ }\href
  {\doibase 10.1103/PhysRevC.91.044312} {\bibfield  {journal} {\bibinfo
  {journal} {Phys. Rev. C}\ }\textbf {\bibinfo {volume} {91}},\ \bibinfo
  {pages} {044312} (\bibinfo {year} {2015}{\natexlab{b}})}\BibitemShut
  {NoStop}%
\bibitem [{\citenamefont {Sword}\ \emph {et~al.}(2019)\citenamefont {Sword},
  \citenamefont {Brett}, \citenamefont {Baumann}, \citenamefont {Brown},
  \citenamefont {Frank}, \citenamefont {Herman}, \citenamefont {Jones},
  \citenamefont {Karrick}, \citenamefont {Kuchera}, \citenamefont
  {Thoennessen}, \citenamefont {Tostevin}, \citenamefont {Tuttle-Timm},\ and\
  \citenamefont {DeYoung}}]{Sword}%
  \BibitemOpen
  \bibfield  {author} {\bibinfo {author} {\bibfnamefont {C.}~\bibnamefont
  {Sword}}, \bibinfo {author} {\bibfnamefont {J.}~\bibnamefont {Brett}},
  \bibinfo {author} {\bibfnamefont {T.}~\bibnamefont {Baumann}}, \bibinfo
  {author} {\bibfnamefont {B.~A.}\ \bibnamefont {Brown}}, \bibinfo {author}
  {\bibfnamefont {N.}~\bibnamefont {Frank}}, \bibinfo {author} {\bibfnamefont
  {J.}~\bibnamefont {Herman}}, \bibinfo {author} {\bibfnamefont {M.~D.}\
  \bibnamefont {Jones}}, \bibinfo {author} {\bibfnamefont {H.}~\bibnamefont
  {Karrick}}, \bibinfo {author} {\bibfnamefont {A.~N.}\ \bibnamefont
  {Kuchera}}, \bibinfo {author} {\bibfnamefont {M.}~\bibnamefont
  {Thoennessen}}, \bibinfo {author} {\bibfnamefont {J.~A.}\ \bibnamefont
  {Tostevin}}, \bibinfo {author} {\bibfnamefont {M.}~\bibnamefont
  {Tuttle-Timm}}, \ and\ \bibinfo {author} {\bibfnamefont {P.~A.}\ \bibnamefont
  {DeYoung}},\ }\href {\doibase 10.1103/PhysRevC.100.034323} {\bibfield
  {journal} {\bibinfo  {journal} {Phys. Rev. C}\ }\textbf {\bibinfo {volume}
  {100}},\ \bibinfo {pages} {034323} (\bibinfo {year} {2019})}\BibitemShut
  {NoStop}%
\bibitem [{\citenamefont {Agostinelli}\ \emph {et~al.}(2003)\citenamefont
  {Agostinelli}, \citenamefont {Allison}, \citenamefont {Amako}, \citenamefont
  {Apostolakis}, \citenamefont {Araujo}, \citenamefont {Arce}, \citenamefont
  {Asai}, \citenamefont {Axen}, \citenamefont {Banerjee}, \citenamefont
  {Barrand}, \citenamefont {Behner}, \citenamefont {Bellagamba}, \citenamefont
  {Boudreau}, \citenamefont {Broglia}, \citenamefont {Brunengo}, \citenamefont
  {Burkhardt}, \citenamefont {Chauvie}, \citenamefont {Chuma}, \citenamefont
  {Chytracek}, \citenamefont {Cooperman}, \citenamefont {Cosmo}, \citenamefont
  {Degtyarenko}, \citenamefont {Dell'Acqua}, \citenamefont {Depaola},
  \citenamefont {Dietrich}, \citenamefont {Enami}, \citenamefont {Feliciello},
  \citenamefont {Ferguson}, \citenamefont {Fesefeldt}, \citenamefont {Folger},
  \citenamefont {Foppiano}, \citenamefont {Forti}, \citenamefont {Garelli},
  \citenamefont {Giani}, \citenamefont {Giannitrapani}, \citenamefont {Gibin},
  \citenamefont {{Gómez Cadenas}}, \citenamefont {González}, \citenamefont
  {{Gracia Abril}}, \citenamefont {Greeniaus}, \citenamefont {Greiner},
  \citenamefont {Grichine}, \citenamefont {Grossheim}, \citenamefont
  {Guatelli}, \citenamefont {Gumplinger}, \citenamefont {Hamatsu},
  \citenamefont {Hashimoto}, \citenamefont {Hasui}, \citenamefont {Heikkinen},
  \citenamefont {Howard}, \citenamefont {Ivanchenko}, \citenamefont {Johnson},
  \citenamefont {Jones}, \citenamefont {Kallenbach}, \citenamefont {Kanaya},
  \citenamefont {Kawabata}, \citenamefont {Kawabata}, \citenamefont {Kawaguti},
  \citenamefont {Kelner}, \citenamefont {Kent}, \citenamefont {Kimura},
  \citenamefont {Kodama}, \citenamefont {Kokoulin}, \citenamefont {Kossov},
  \citenamefont {Kurashige}, \citenamefont {Lamanna}, \citenamefont {Lampén},
  \citenamefont {Lara}, \citenamefont {Lefebure}, \citenamefont {Lei},
  \citenamefont {Liendl}, \citenamefont {Lockman}, \citenamefont {Longo},
  \citenamefont {Magni}, \citenamefont {Maire}, \citenamefont {Medernach},
  \citenamefont {Minamimoto}, \citenamefont {{Mora de Freitas}}, \citenamefont
  {Morita}, \citenamefont {Murakami}, \citenamefont {Nagamatu}, \citenamefont
  {Nartallo}, \citenamefont {Nieminen}, \citenamefont {Nishimura},
  \citenamefont {Ohtsubo}, \citenamefont {Okamura}, \citenamefont {O'Neale},
  \citenamefont {Oohata}, \citenamefont {Paech}, \citenamefont {Perl},
  \citenamefont {Pfeiffer}, \citenamefont {Pia}, \citenamefont {Ranjard},
  \citenamefont {Rybin}, \citenamefont {Sadilov}, \citenamefont {{Di Salvo}},
  \citenamefont {Santin}, \citenamefont {Sasaki}, \citenamefont {Savvas},
  \citenamefont {Sawada}, \citenamefont {Scherer}, \citenamefont {Sei},
  \citenamefont {Sirotenko}, \citenamefont {Smith}, \citenamefont {Starkov},
  \citenamefont {Stoecker}, \citenamefont {Sulkimo}, \citenamefont {Takahata},
  \citenamefont {Tanaka}, \citenamefont {Tcherniaev}, \citenamefont {{Safai
  Tehrani}}, \citenamefont {Tropeano}, \citenamefont {Truscott}, \citenamefont
  {Uno}, \citenamefont {Urban}, \citenamefont {Urban}, \citenamefont {Verderi},
  \citenamefont {Walkden}, \citenamefont {Wander}, \citenamefont {Weber},
  \citenamefont {Wellisch}, \citenamefont {Wenaus}, \citenamefont {Williams},
  \citenamefont {Wright}, \citenamefont {Yamada}, \citenamefont {Yoshida},\
  and\ \citenamefont {Zschiesche}}]{Geant4}%
  \BibitemOpen
  \bibfield  {author} {\bibinfo {author} {\bibfnamefont {S.}~\bibnamefont
  {Agostinelli}}, \bibinfo {author} {\bibfnamefont {J.}~\bibnamefont
  {Allison}}, \bibinfo {author} {\bibfnamefont {K.}~\bibnamefont {Amako}},
  \bibinfo {author} {\bibfnamefont {J.}~\bibnamefont {Apostolakis}}, \bibinfo
  {author} {\bibfnamefont {H.}~\bibnamefont {Araujo}}, \bibinfo {author}
  {\bibfnamefont {P.}~\bibnamefont {Arce}}, \bibinfo {author} {\bibfnamefont
  {M.}~\bibnamefont {Asai}}, \bibinfo {author} {\bibfnamefont {D.}~\bibnamefont
  {Axen}}, \bibinfo {author} {\bibfnamefont {S.}~\bibnamefont {Banerjee}},
  \bibinfo {author} {\bibfnamefont {G.}~\bibnamefont {Barrand}}, \bibinfo
  {author} {\bibfnamefont {F.}~\bibnamefont {Behner}}, \bibinfo {author}
  {\bibfnamefont {L.}~\bibnamefont {Bellagamba}}, \bibinfo {author}
  {\bibfnamefont {J.}~\bibnamefont {Boudreau}}, \bibinfo {author}
  {\bibfnamefont {L.}~\bibnamefont {Broglia}}, \bibinfo {author} {\bibfnamefont
  {A.}~\bibnamefont {Brunengo}}, \bibinfo {author} {\bibfnamefont
  {H.}~\bibnamefont {Burkhardt}}, \bibinfo {author} {\bibfnamefont
  {S.}~\bibnamefont {Chauvie}}, \bibinfo {author} {\bibfnamefont
  {J.}~\bibnamefont {Chuma}}, \bibinfo {author} {\bibfnamefont
  {R.}~\bibnamefont {Chytracek}}, \bibinfo {author} {\bibfnamefont
  {G.}~\bibnamefont {Cooperman}}, \bibinfo {author} {\bibfnamefont
  {G.}~\bibnamefont {Cosmo}}, \bibinfo {author} {\bibfnamefont
  {P.}~\bibnamefont {Degtyarenko}}, \bibinfo {author} {\bibfnamefont
  {A.}~\bibnamefont {Dell'Acqua}}, \bibinfo {author} {\bibfnamefont
  {G.}~\bibnamefont {Depaola}}, \bibinfo {author} {\bibfnamefont
  {D.}~\bibnamefont {Dietrich}}, \bibinfo {author} {\bibfnamefont
  {R.}~\bibnamefont {Enami}}, \bibinfo {author} {\bibfnamefont
  {A.}~\bibnamefont {Feliciello}}, \bibinfo {author} {\bibfnamefont
  {C.}~\bibnamefont {Ferguson}}, \bibinfo {author} {\bibfnamefont
  {H.}~\bibnamefont {Fesefeldt}}, \bibinfo {author} {\bibfnamefont
  {G.}~\bibnamefont {Folger}}, \bibinfo {author} {\bibfnamefont
  {F.}~\bibnamefont {Foppiano}}, \bibinfo {author} {\bibfnamefont
  {A.}~\bibnamefont {Forti}}, \bibinfo {author} {\bibfnamefont
  {S.}~\bibnamefont {Garelli}}, \bibinfo {author} {\bibfnamefont
  {S.}~\bibnamefont {Giani}}, \bibinfo {author} {\bibfnamefont
  {R.}~\bibnamefont {Giannitrapani}}, \bibinfo {author} {\bibfnamefont
  {D.}~\bibnamefont {Gibin}}, \bibinfo {author} {\bibfnamefont
  {J.}~\bibnamefont {{Gómez Cadenas}}}, \bibinfo {author} {\bibfnamefont
  {I.}~\bibnamefont {González}}, \bibinfo {author} {\bibfnamefont
  {G.}~\bibnamefont {{Gracia Abril}}}, \bibinfo {author} {\bibfnamefont
  {G.}~\bibnamefont {Greeniaus}}, \bibinfo {author} {\bibfnamefont
  {W.}~\bibnamefont {Greiner}}, \bibinfo {author} {\bibfnamefont
  {V.}~\bibnamefont {Grichine}}, \bibinfo {author} {\bibfnamefont
  {A.}~\bibnamefont {Grossheim}}, \bibinfo {author} {\bibfnamefont
  {S.}~\bibnamefont {Guatelli}}, \bibinfo {author} {\bibfnamefont
  {P.}~\bibnamefont {Gumplinger}}, \bibinfo {author} {\bibfnamefont
  {R.}~\bibnamefont {Hamatsu}}, \bibinfo {author} {\bibfnamefont
  {K.}~\bibnamefont {Hashimoto}}, \bibinfo {author} {\bibfnamefont
  {H.}~\bibnamefont {Hasui}}, \bibinfo {author} {\bibfnamefont
  {A.}~\bibnamefont {Heikkinen}}, \bibinfo {author} {\bibfnamefont
  {A.}~\bibnamefont {Howard}}, \bibinfo {author} {\bibfnamefont
  {V.}~\bibnamefont {Ivanchenko}}, \bibinfo {author} {\bibfnamefont
  {A.}~\bibnamefont {Johnson}}, \bibinfo {author} {\bibfnamefont
  {F.}~\bibnamefont {Jones}}, \bibinfo {author} {\bibfnamefont
  {J.}~\bibnamefont {Kallenbach}}, \bibinfo {author} {\bibfnamefont
  {N.}~\bibnamefont {Kanaya}}, \bibinfo {author} {\bibfnamefont
  {M.}~\bibnamefont {Kawabata}}, \bibinfo {author} {\bibfnamefont
  {Y.}~\bibnamefont {Kawabata}}, \bibinfo {author} {\bibfnamefont
  {M.}~\bibnamefont {Kawaguti}}, \bibinfo {author} {\bibfnamefont
  {S.}~\bibnamefont {Kelner}}, \bibinfo {author} {\bibfnamefont
  {P.}~\bibnamefont {Kent}}, \bibinfo {author} {\bibfnamefont {A.}~\bibnamefont
  {Kimura}}, \bibinfo {author} {\bibfnamefont {T.}~\bibnamefont {Kodama}},
  \bibinfo {author} {\bibfnamefont {R.}~\bibnamefont {Kokoulin}}, \bibinfo
  {author} {\bibfnamefont {M.}~\bibnamefont {Kossov}}, \bibinfo {author}
  {\bibfnamefont {H.}~\bibnamefont {Kurashige}}, \bibinfo {author}
  {\bibfnamefont {E.}~\bibnamefont {Lamanna}}, \bibinfo {author} {\bibfnamefont
  {T.}~\bibnamefont {Lampén}}, \bibinfo {author} {\bibfnamefont
  {V.}~\bibnamefont {Lara}}, \bibinfo {author} {\bibfnamefont {V.}~\bibnamefont
  {Lefebure}}, \bibinfo {author} {\bibfnamefont {F.}~\bibnamefont {Lei}},
  \bibinfo {author} {\bibfnamefont {M.}~\bibnamefont {Liendl}}, \bibinfo
  {author} {\bibfnamefont {W.}~\bibnamefont {Lockman}}, \bibinfo {author}
  {\bibfnamefont {F.}~\bibnamefont {Longo}}, \bibinfo {author} {\bibfnamefont
  {S.}~\bibnamefont {Magni}}, \bibinfo {author} {\bibfnamefont
  {M.}~\bibnamefont {Maire}}, \bibinfo {author} {\bibfnamefont
  {E.}~\bibnamefont {Medernach}}, \bibinfo {author} {\bibfnamefont
  {K.}~\bibnamefont {Minamimoto}}, \bibinfo {author} {\bibfnamefont
  {P.}~\bibnamefont {{Mora de Freitas}}}, \bibinfo {author} {\bibfnamefont
  {Y.}~\bibnamefont {Morita}}, \bibinfo {author} {\bibfnamefont
  {K.}~\bibnamefont {Murakami}}, \bibinfo {author} {\bibfnamefont
  {M.}~\bibnamefont {Nagamatu}}, \bibinfo {author} {\bibfnamefont
  {R.}~\bibnamefont {Nartallo}}, \bibinfo {author} {\bibfnamefont
  {P.}~\bibnamefont {Nieminen}}, \bibinfo {author} {\bibfnamefont
  {T.}~\bibnamefont {Nishimura}}, \bibinfo {author} {\bibfnamefont
  {K.}~\bibnamefont {Ohtsubo}}, \bibinfo {author} {\bibfnamefont
  {M.}~\bibnamefont {Okamura}}, \bibinfo {author} {\bibfnamefont
  {S.}~\bibnamefont {O'Neale}}, \bibinfo {author} {\bibfnamefont
  {Y.}~\bibnamefont {Oohata}}, \bibinfo {author} {\bibfnamefont
  {K.}~\bibnamefont {Paech}}, \bibinfo {author} {\bibfnamefont
  {J.}~\bibnamefont {Perl}}, \bibinfo {author} {\bibfnamefont {A.}~\bibnamefont
  {Pfeiffer}}, \bibinfo {author} {\bibfnamefont {M.}~\bibnamefont {Pia}},
  \bibinfo {author} {\bibfnamefont {F.}~\bibnamefont {Ranjard}}, \bibinfo
  {author} {\bibfnamefont {A.}~\bibnamefont {Rybin}}, \bibinfo {author}
  {\bibfnamefont {S.}~\bibnamefont {Sadilov}}, \bibinfo {author} {\bibfnamefont
  {E.}~\bibnamefont {{Di Salvo}}}, \bibinfo {author} {\bibfnamefont
  {G.}~\bibnamefont {Santin}}, \bibinfo {author} {\bibfnamefont
  {T.}~\bibnamefont {Sasaki}}, \bibinfo {author} {\bibfnamefont
  {N.}~\bibnamefont {Savvas}}, \bibinfo {author} {\bibfnamefont
  {Y.}~\bibnamefont {Sawada}}, \bibinfo {author} {\bibfnamefont
  {S.}~\bibnamefont {Scherer}}, \bibinfo {author} {\bibfnamefont
  {S.}~\bibnamefont {Sei}}, \bibinfo {author} {\bibfnamefont {V.}~\bibnamefont
  {Sirotenko}}, \bibinfo {author} {\bibfnamefont {D.}~\bibnamefont {Smith}},
  \bibinfo {author} {\bibfnamefont {N.}~\bibnamefont {Starkov}}, \bibinfo
  {author} {\bibfnamefont {H.}~\bibnamefont {Stoecker}}, \bibinfo {author}
  {\bibfnamefont {J.}~\bibnamefont {Sulkimo}}, \bibinfo {author} {\bibfnamefont
  {M.}~\bibnamefont {Takahata}}, \bibinfo {author} {\bibfnamefont
  {S.}~\bibnamefont {Tanaka}}, \bibinfo {author} {\bibfnamefont
  {E.}~\bibnamefont {Tcherniaev}}, \bibinfo {author} {\bibfnamefont
  {E.}~\bibnamefont {{Safai Tehrani}}}, \bibinfo {author} {\bibfnamefont
  {M.}~\bibnamefont {Tropeano}}, \bibinfo {author} {\bibfnamefont
  {P.}~\bibnamefont {Truscott}}, \bibinfo {author} {\bibfnamefont
  {H.}~\bibnamefont {Uno}}, \bibinfo {author} {\bibfnamefont {L.}~\bibnamefont
  {Urban}}, \bibinfo {author} {\bibfnamefont {P.}~\bibnamefont {Urban}},
  \bibinfo {author} {\bibfnamefont {M.}~\bibnamefont {Verderi}}, \bibinfo
  {author} {\bibfnamefont {A.}~\bibnamefont {Walkden}}, \bibinfo {author}
  {\bibfnamefont {W.}~\bibnamefont {Wander}}, \bibinfo {author} {\bibfnamefont
  {H.}~\bibnamefont {Weber}}, \bibinfo {author} {\bibfnamefont
  {J.}~\bibnamefont {Wellisch}}, \bibinfo {author} {\bibfnamefont
  {T.}~\bibnamefont {Wenaus}}, \bibinfo {author} {\bibfnamefont
  {D.}~\bibnamefont {Williams}}, \bibinfo {author} {\bibfnamefont
  {D.}~\bibnamefont {Wright}}, \bibinfo {author} {\bibfnamefont
  {T.}~\bibnamefont {Yamada}}, \bibinfo {author} {\bibfnamefont
  {H.}~\bibnamefont {Yoshida}}, \ and\ \bibinfo {author} {\bibfnamefont
  {D.}~\bibnamefont {Zschiesche}},\ }\href {\doibase
  https://doi.org/10.1016/S0168-9002(03)01368-8} {\bibfield  {journal}
  {\bibinfo  {journal} {Nuclear Instruments and Methods in Physics Research
  Section A: Accelerators, Spectrometers, Detectors and Associated Equipment}\
  }\textbf {\bibinfo {volume} {506}},\ \bibinfo {pages} {250} (\bibinfo {year}
  {2003})}\BibitemShut {NoStop}%
\bibitem [{\citenamefont {Roeder}(2008)}]{Menate}%
  \BibitemOpen
  \bibfield  {author} {\bibinfo {author} {\bibfnamefont {B.}~\bibnamefont
  {Roeder}},\ }\href@noop {} {\bibfield  {journal} {\bibinfo  {journal}
  {EURISOL Design Study}\ }\textbf {\bibinfo {volume} {Report No.
  10-25-2008-006-In-beamvalidations.pdf}},\ \bibinfo {pages} {p. 31} (\bibinfo
  {year} {2008})}\BibitemShut {NoStop}%
\bibitem [{\citenamefont {Jones}\ \emph {et~al.}(2017)\citenamefont {Jones},
  \citenamefont {Fossez}, \citenamefont {Baumann}, \citenamefont {DeYoung},
  \citenamefont {Finck}, \citenamefont {Frank}, \citenamefont {Kuchera},
  \citenamefont {Michel}, \citenamefont {Nazarewicz}, \citenamefont {Rotureau},
  \citenamefont {Smith}, \citenamefont {Stephenson}, \citenamefont {Stiefel},
  \citenamefont {Thoennessen},\ and\ \citenamefont {Zegers}}]{Jones2017}%
  \BibitemOpen
  \bibfield  {author} {\bibinfo {author} {\bibfnamefont {M.~D.}\ \bibnamefont
  {Jones}}, \bibinfo {author} {\bibfnamefont {K.}~\bibnamefont {Fossez}},
  \bibinfo {author} {\bibfnamefont {T.}~\bibnamefont {Baumann}}, \bibinfo
  {author} {\bibfnamefont {P.~A.}\ \bibnamefont {DeYoung}}, \bibinfo {author}
  {\bibfnamefont {J.~E.}\ \bibnamefont {Finck}}, \bibinfo {author}
  {\bibfnamefont {N.}~\bibnamefont {Frank}}, \bibinfo {author} {\bibfnamefont
  {A.~N.}\ \bibnamefont {Kuchera}}, \bibinfo {author} {\bibfnamefont
  {N.}~\bibnamefont {Michel}}, \bibinfo {author} {\bibfnamefont
  {W.}~\bibnamefont {Nazarewicz}}, \bibinfo {author} {\bibfnamefont
  {J.}~\bibnamefont {Rotureau}}, \bibinfo {author} {\bibfnamefont {J.~K.}\
  \bibnamefont {Smith}}, \bibinfo {author} {\bibfnamefont {S.~L.}\ \bibnamefont
  {Stephenson}}, \bibinfo {author} {\bibfnamefont {K.}~\bibnamefont {Stiefel}},
  \bibinfo {author} {\bibfnamefont {M.}~\bibnamefont {Thoennessen}}, \ and\
  \bibinfo {author} {\bibfnamefont {R.~G.~T.}\ \bibnamefont {Zegers}},\ }\href
  {\doibase 10.1103/PhysRevC.96.054322} {\bibfield  {journal} {\bibinfo
  {journal} {Phys. Rev. C}\ }\textbf {\bibinfo {volume} {96}},\ \bibinfo
  {pages} {054322} (\bibinfo {year} {2017})}\BibitemShut {NoStop}%
\bibitem [{\citenamefont {Wang}\ \emph {et~al.}(2021)\citenamefont {Wang},
  \citenamefont {Huang}, \citenamefont {Kondev}, \citenamefont {Audi},\ and\
  \citenamefont {Naimi}}]{AME20}%
  \BibitemOpen
  \bibfield  {author} {\bibinfo {author} {\bibfnamefont {M.}~\bibnamefont
  {Wang}}, \bibinfo {author} {\bibfnamefont {W.}~\bibnamefont {Huang}},
  \bibinfo {author} {\bibfnamefont {F.}~\bibnamefont {Kondev}}, \bibinfo
  {author} {\bibfnamefont {G.}~\bibnamefont {Audi}}, \ and\ \bibinfo {author}
  {\bibfnamefont {S.}~\bibnamefont {Naimi}},\ }\href {\doibase
  10.1088/1674-1137/abddaf} {\bibfield  {journal} {\bibinfo  {journal} {Chinese
  Physics C}\ }\textbf {\bibinfo {volume} {45}},\ \bibinfo {pages} {030003}
  (\bibinfo {year} {2021})}\BibitemShut {NoStop}%
\end{thebibliography}%

\end{document}